# Mechanisms of charge transfer and redistribution in LaAlO$_3$/SrTiO$_3$ revealed by high-energy optical conductivity

T. C. Asmara[1,2,3], A. Annadi[1,2], I. Santoso[1,2,3], P. K. Gogoi[1,2,3], A. Kotlov[4], H. M. Omer[3], M. Motapothula[1,2,7], M. B. H. Breese[2,3,7], M. Rübhausen[1,6], T. Venkatesan[1,2,5], Ariando[1,2], A. Rusydi[1,2,3*]

[1] *NUSNNI-NanoCore, National University of Singapore, Singapore 117411.*

[2] *Department of Physics, National University of Singapore, Singapore 117576.*

[3] *Singapore Synchrotron Light Source, National University of Singapore, Singapore 117603.*

[4] *Hamburger Synchrotronstrahlungslabor (HASYLAB) at Deutsches Elektronen-Synchrotron (DESY), Notkestrasse 85, 22603 Hamburg, Germany.*

[5] *Department of Electrical and Computer Engineering, National University of Singapore, Singapore 117583.*

[6] *Institut für Angewandte Physik, Universität Hamburg, Jungiusstrasse 11, D-20355 Hamburg, Germany. Center for Free Electron Laser Science (CFEL), Notkesstrasse 85, D-22607 Hamburg, Germany.*

[7] *Center for Ion Beam Applications (CIBA), Department of Physics, National University of Singapore, Singapore 117542.*

*Correspondence to: phyandri@nus.edu.sg

**In condensed matter physics the quasi two-dimensional electron gas at the interface of two different insulators, polar LaAlO$_3$ on non-polar SrTiO$_3$ (LaAlO$_3$/SrTiO$_3$) is a spectacular and surprising observation. This phenomenon is LaAlO$_3$ film thickness-dependent and may be explained by the polarization catastrophe model, in which a charge transfer of 0.5$e^-$ from the LaAlO$_3$ film into the LaAlO$_3$/SrTiO$_3$ interface is expected. Here we show that in conducting samples (≥4 unit cells of LaAlO$_3$) there is indeed a ~0.5$e^-$ transfer from LaAlO$_3$ into the LaAlO$_3$/SrTiO$_3$ interface by studying the optical conductivity in a broad energy range (0.5–35 eV). Surprisingly, in insulating samples (≤3 unit cells of LaAlO$_3$) a redistribution of charges within the polar LaAlO$_3$ sub-layers (from AlO$_2$ to LaO) as large as ~0.5$e^-$ is observed, with no charge transfer into the interface. Hence, our results reveal the different mechanisms for the polarization catastrophe compensation in insulating and conducting LaAlO$_3$/SrTiO$_3$ interfaces.**

Some of the most exciting condensed matter physics problems are found at the interfaces of dissimilar materials[1]. The behaviour of electrons at these interfaces would be governed by electronic reconstruction mechanisms[2] leading to a variety of exotic quantum phenomena[1]. In conjunction with x-ray and electron spectroscopy techniques[3-7] with their inherent advantages and constraints, an experimental technique that can directly reveal hidden quantum phenomena at buried interfaces is highly desirable. In this paper we demonstrate the potency of high-energy optical reflectivity coupled with spectroscopic ellipsometry and study an interface consisting of two dissimilar insulators: polar LaAlO$_3$ and non-polar SrTiO$_3$ revealing the details of the charge (electron) transfer among and within the layers that govern the conductivity of the buried interface.

The quasi two-dimensional electron gas (2DEG) at the the buried interface of two different insulator oxides heterostructure, polar LaAlO$_3$ on non-polar SrTiO$_3$ (LaAlO$_3$/SrTiO$_3$)[8] has shown many interesting phenomena ranging from metal-insulator transition[9], superconductivity[10], and magnetism[11-14]. According to the controversial but compelling polarization catastrophe model, the polar sub-layers of LaAlO$_3$ ((LaO)$^+$ and (AlO$_2$)$^-$) give rise to a polarization field inside LaAlO$_3$ that causes an electronic potential build-up as the LaAlO$_3$ film thickness increases. To counter this, a charge transfer of 0.5$e^-$ per unit cell (uc) (~3×10$^{14}$ cm$^{-2}$) from LaAlO$_3$ into the LaAlO$_3$/SrTiO$_3$ interface is required[15,16]. Various techniques have shown a charge transfer much less than this. For example, x-ray-based techniques[5,6] have estimated up to 1.1×10$^{14}$ cm$^{-2}$ while transport measurements[9-11] yield substantially smaller number of carriers of ~10$^{13}$ cm$^{-2}$. It has been suggested that charge localization effects might limit the number of mobile charges that can be measured by transport[7,17], and thus if a technique can measure and quantify both the localized and delocalized charges, one might be able to evaluate the actual charge transfer[5-7].

Another unresolved important issue is the insulating case of LaAlO$_3$/SrTiO$_3$ (≤3 uc of LaAlO$_3$). Transport measurements[9] have shown that the conducting interface only exists above a certain critical thickness of LaAlO$_3$, typically ≥4 uc (although cationic stoichiometry, *e.g.*, the La/Al ratio of LaAlO$_3$ film, may also affect the interface conductivity, with conducting LaAlO$_3$/SrTiO$_3$ observed to have slightly Al-rich LaAlO$_3$ film[18]). This means that the charge transfer into the interface required for countering the polarization catastrophe does not happen when the thickness of LaAlO$_3$ is below 4 uc. According to the prevalent polarization catastrophe model, this means the polarization field should be present for ≤3 uc of LaAlO$_3$. One way to verify the model is to measure this polarization potential build-up in insulating LaAlO$_3$/SrTiO$_3$, which is predicted to be 0.24 V/Å (or ~0.9 V per uc of LaAlO$_3$)[19]. However, attempts to measure this have not been successful using core-level x-ray photoemission spectroscopy (XPS)[20,21], in which the measured core-level shift in LaAlO$_3$ is only ~0.1 eV per uc of LaAlO$_3$, much less than expected. If the changes in the band structure are predominantly near the valence bands and the Fermi level, then the appropriate technique should directly probe states near the valence bands and the Fermi level.

Let us approach the problem from a different angle. Another way to overcome the polarization potential is by charge redistribution within the LaAlO$_3$ layers. In a *Gedankenexperiment*, we hypothesize an extreme case of charge redistribution of 1$e^-$ between the AlO$_2$ and LaO sub-layers of LaAlO$_3$, which is also adequate to compensate the polarization potential, although the actual amount of charge redistribution might be restricted by electrostatics. Hence, instead of measuring the potential build-up in the layers, one can measure the charge redistribution within the layers directly. This can be done by measuring the optical conductivity involving states below and above the valence bands, the conduction bands, and the Fermi level, and then use the f-sum rule, which represents charge conservation, to quantify the charge redistribution.

Furthermore, recent band structure calculations and surface x-ray diffraction measurements suggest that distortions of the LaAlO$_3$ lattice (buckling) may partly compensate the polarization field in insulating LaAlO$_3$/SrTiO$_3$[22-25]. Interestingly, when the interface becomes conducting at $\geq$ 4 uc of LaAlO$_3$, this distortion decreases and ultimately vanishes[24,25], indicating that the buckling mechanism is unique to the insulating case of LaAlO$_3$/SrTiO$_3$. This raises another important question: since the buckling is a structural change, will the electronic structure change appropriately, and manifest as a charge redistribution within the LaAlO$_3$ layer itself? Thus, it is again critical to be able to measure these intra-layer charge redistributions.

As mentioned earlier, a direct way to probe the electronic band structure and charge (localized and delocalized) redistribution mechanisms is to measure the complex dielectric response of the material, from which the optical conductivity can be extracted in a broad energy range[26-28]. Here, we use a combination of spectroscopic ellipsometry and ultraviolet – vacuum ultraviolet (UV-VUV) reflectivity to probe the intrinsic properties of the LaAlO$_3$/SrTiO$_3$ interface using photon with energies between 0.5–35 eV. Due to a stabilized Kramers-Kronig transformation[29,30], the strength of this experimental approach allows one to measure the charge transfer of both delocalized and localized charges accurately using the optical f-sum rule. Since localized electrons are inaccessible to electrical transport measurements, but are accessible by photons, we overcome a severe constraint. In particular, the optical transitions involving AlO$_2$ sub-layer of LaAlO$_3$ is very distinct and well-separated from the ones involving LaO sub-layer, so that the internal charge redistribution within the LaAlO$_3$ sub-layers can be clearly identified. The same is also true for the TiO$_2$ and SrO sub-layers of SrTiO$_3$.

Here, we show that in conducting LaAlO$_3$/SrTiO$_3$ (4 and 6 uc of LaAlO$_3$ film on SrTiO$_3$) there is indeed a charge transfer from LaAlO$_3$ into the interface, and that the amount of charge transfer is ~0.5$e^-$. In the insulating case (2 and 3 uc of LaAlO$_3$ film on SrTiO$_3$), we surprisingly observe ~0.5$e^-$ charge redistribution from AlO$_2$ to LaO sub-layers, within the LaAlO$_3$ layers. This suggests that for the insulating case the polarization catastrophe could be partly overcome by the above-mentioned charge redistribution, which may be a consequence of the buckling of the LaAlO$_3$ lattice.

**Results**

**Structural and transport measurements.** LaAlO$_3$/SrTiO$_3$ samples were prepared by growing LaAlO$_3$ film on top of TiO$_2$-terminated (001) SrTiO$_3$ using pulsed-laser deposition (PLD)[12]. The atomic force microscopy (AFM) topography image of the TiO$_2$ terminated SrTiO$_3$ substrate in Fig. 1a clearly shows the atomically flat surface with unit cell steps. Four high-quality samples with varying thickness of 2, 3, 4, and 6 uc of LaAlO$_3$ film were prepared as a model interface system for the high-energy optical studies. The growth of the films was monitored using reflective high-energy electron diffraction (RHEED) (Fig.1b). After LaAlO$_3$ deposition,

AFM topography measurements show that the atomically flat surface with unit cell step and terrace structure of SrTiO$_3$ is preserved, with surface roughness of ~1 Å (see Figs. 1c and 1d). This ensures that surface roughness effects do not adversely affect the optical measurements. Transport measurements (Fig. 1e), which were taken before and after the optics measurements, show consistently that 2 and 3 uc samples are insulating with carrier density and conductivity below the measurement limit, while 4 and 6 uc ones are conducting with carrier density of 4-6×10$^{13}$ cm$^{-2}$ and conductivity of 4-8×10$^{-5}$ Ω$^{-1}$, consistent with previous transport results[9-11].

The perovskite LaAlO$_3$ unit cell (Fig. 1f) can be divided into two sub-layers: LaO and AlO$_2$, in which theoretical calculations have shown that the band structures of these discriminated sub-layers are indeed different[19,22], leading to distinct optical transitions. To accommodate the assignments of these optical transitions, we define O$_{La}$ as the O in the LaO plane and O$_{Al}$ as the O in the AlO$_2$ plane. Likewise, SrTiO$_3$ also has similar layered perovskite structure and thus can also be divided into two sub-layers: SrO and TiO$_2$. Then O$_{Sr}$ is defined as O that belongs to SrO sub-layer, while O$_{Ti}$ is defined as the one in TiO$_2$. Similarly, the O in the different planes of SrTiO$_3$ can also lead to distinct optical transitions. This discrimination is important, as discussed later, because it can reveal the intra- and inter-layer charge transfer mechanism in LaAlO$_3$/SrTiO$_3$ for both the insulating and conducting samples.

**Spectroscopic ellipsometry and high-energy reflectivity.** Our main observation is the high-energy reflectivity of LaAlO$_3$/SrTiO$_3$ at different thicknesses of LaAlO$_3$ as compared to bulk LaAlO$_3$ and SrTiO$_3$ as shown in Fig. 2a. Note that, due to the challenge in making optical measurements over such a broad energy range, in this study we have only measured a selected set of samples as representative of insulating (2 and 3 uc) and conducting (4 and 6 uc) LaAlO$_3$/SrTiO$_3$. Thus, further measurements on a larger set of samples may be important in further deepening our analyses. It can be seen that the reflectivities of the insulating 2 and 3 uc

LaAlO$_3$/SrTiO$_3$ are similar, and the same is true for the conducting 4 and 6 uc LaAlO$_3$/SrTiO$_3$. Surprisingly, there are huge differences between reflectivity of insulating and conducting samples. These differences occur more significantly at high photon energies, particularly in the energy ranges of 9–14 eV and 14–21 eV. In the 9–14 eV range, the reflectivity of conducting samples is lower than insulating samples, while in the 14–21 eV range the opposite occurs. In contrast, between 4–9 eV, the differences are less, and below 4 eV negligible. This signifies why going beyond conventional (up to ~5 eV) spectroscopic ellipsometry is important. (Note that the spectroscopic ellipsometry data is crucial for the normalization of the derived dielectric functions from the reflectivity measurements made up to 35 eV as shown in Supplementary Methods.) The electronic band structures of the insulating and conducting LaAlO$_3$/SrTiO$_3$ are very different at high energy, and these differences are critical in revealing the true nature of LaAlO$_3$/SrTiO$_3$ interface. Furthermore, since reflectivity and spectroscopic ellipsometry are sensitive to unpercolated clusters of charges[31], the similarity of the reflectivity of insulating 2 and 3 uc LaAlO$_3$/SrTiO$_3$ also implies that there is no evidence of precursor of percolation effects in the insulating samples, especially the 3 uc LaAlO$_3$/SrTiO$_3$[32].

**Discussion**

For detailed analysis, we turn our discussion to optical conductivity, $\sigma_1$, because it fulfils the optical f-sum rule, which is related to number of charges excited by the photons. For bulk materials like bulk LaAlO$_3$ and bulk SrTiO$_3$, $\sigma_1$ can be extracted directly from reflectivity using Kramers-Kronig analysis[29,30]. On the other hand, LaAlO$_3$/SrTiO$_3$ is layered along the <001> direction (perpendicular to the (001) surface of the sample) due to its heterostructure nature as well as the presence of the conducting layer at LaAlO$_3$/SrTiO$_3$ interface. For this reason, the reflectivity of LaAlO$_3$/SrTiO$_3$ is analyzed based on standard theory of wave propagation in a stratified media[33,34]. The analysis naturally leads to a three-layered structure for the conducting LaAlO$_3$/SrTiO$_3$: LaAlO$_3$ film layer on top, bulk SrTiO$_3$ substrate at the bottom, and an interface layer sandwiched in between, representing the 2DEG of the conducting samples.

A self-consistent iteration procedure is used to extract the thickness and dielectric function of each layer, and as long as the iteration is convergent, the starting assumption of these parameters should have little effect, if any, on the final obtained values (see Supplementary Methods for details). From the analysis, it is found that the thickness of this interface layer is ~5 nm, consistent with previous observation using hard XPS[5], cross-sectional conducting tip atomic force microscopy[35], and the upper limit for the superconducting layer thickness of LaAlO$_3$/SrTiO$_3$[10]. This result also suggests that the high-energy reflectivity can be used to measure the thickness of interface layer. For insulating LaAlO$_3$/SrTiO$_3$, the analysis naturally converges into an effective two-layered structure instead. This means for insulating LaAlO$_3$/SrTiO$_3$ the $\sigma_1$ at interface is very similar to that of bulk SrTiO$_3$ as discussed later. This can be easily understood due to the absence of the conducting interface layer.

Now, $\sigma_1$ of each individual layer can be extracted separately, so that we can analyze the concomitant evolution of each individual layer of LaAlO$_3$/SrTiO$_3$ as the interface changes from insulating to conducting. Spectra of $\sigma_1$ for each layer are shown in Figs. 2b and 2c. It should be noted that the plots for 2 and 3 unit cells (uc) LaAlO$_3$/SrTiO$_3$ are the same due to the nature of the iteration process (see Supplementary Methods), and the same is true for the 4 and 6 uc LaAlO$_3$/SrTiO$_3$.

For LaAlO$_3$, one can, based on band structure calculations[36,37], divide $\sigma_1$ into three main optical regions, while $\sigma_1$ of SrTiO$_3$ can be divided into five main optical regions[38-40]. Every transition is unique and originates from different orbitals in each layer and sub-layer, and these are summarized in Table 1. Furthermore, the polarization of the incident light is also taken into account in assigning the optical transitions. Since the incident light is linearly-polarized parallel to the sample surface, the majority of the optical transitions occur in the in-plane direction within each sub-layer, allowing us to study spectral weight transfers between the different sub-layers. For example, in $A_1$ region of LaAlO$_3$ the transition is from $O_{La}$-2p to La-4d, 5f, both of which reside within the LaO sub-layer of LaAlO$_3$. The other transitions also follow this convention.

Fig. 2b shows that $\sigma_1$ of LaAlO$_3$ film of insulating and conducting LaAlO$_3$/SrTiO$_3$ is dramatically different as compared to bulk LaAlO$_3$. Particularly, $\sigma_1$ in $A_1$ region of LaAlO$_3$ film in the insulating samples is higher than the bulk value, while for the conducting samples it is lower. Meanwhile, the reverse is true in $A_2$ region. It is very clear that there are spectral weight transfers occurring between these three regions when the thickness of LaAlO$_3$ film increases and the interface goes from insulating to conducting state.

It can be seen (Fig. 2c) that $\sigma_1$ of the interface layer to a significant extent resembles $\sigma_1$ of the bulk SrTiO$_3$. This indicates that the electronic interface layer is SrTiO$_3$-like, and that the conducting layer mostly resides in SrTiO$_3$-side rather than LaAlO$_3$. The most significant change in $\sigma_1$ happens at $B_3$ region when the interface becomes conducting. In bulk SrTiO$_3$, that region corresponds to a valley with no main optical transition. Interestingly for the conducting samples a completely new peak emerges in that region. This implies that when the interface becomes conducting, a new characteristic interface state emerges representing the presence of the 2DEG. According to previous reports[3-7,17,19,22], the 2DEG resides in the Ti-3d-$t_{2g}$ state of SrTiO$_3$, so this new interface state should also have Ti-3d-$t_{2g}$ characteristic. Thus, based on the optical selection rules, the optical transition at $B_3$ region may be assigned to originate from this new interface state to unoccupied states of higher O orbitals (see Table 1). One should note that the $\sigma_1$ spectra of the interface layer of conducting LaAlO$_3$/SrTiO$_3$ does not show Drude response, consistent with previous infrared spectroscopic ellipsometry experiment[41].

The $\sigma_1$ analysis is very important because it can be linked to the effective number of electrons associated with a particular optical transition, $N$, using partial f-sum rule,

$$\frac{N}{V} = \frac{4m}{he^2} \int_{E_1}^{E_2} \sigma_1(E) dE, \quad (1)$$

where $e$ is the elementary charge, $m$ is the electron mass and $V$ is the unit volume. The $E_1$ and $E_2$ indicate the energy boundaries of that particular transition in the $\sigma_1$ plot. We then define $n_{eff}$ as the $N$ of each layer relative to either bulk LaAlO$_3$ (for LaAlO$_3$ film) or bulk SrTiO$_3$ (for interface layer) values. The advantage of this definition is that any changes in $N$ in LaAlO$_3$ film or the interface layer can be distinguished from the bulk properties.

In LaAlO$_3$/SrTiO$_3$ (and thin films in general), the thickness of the LaAlO$_3$ film and interface layer is finite, so $n_{eff}$ distributes over this finite thickness. If we assume that the distribution is uniform over each uc, the $n_{eff}$ per uc, $n_{uc}$, can be defined such that,

$$n_{eff} = \int_0^d n_{uc} dx = n_{uc} d, \quad (2)$$

where $d$ is the thickness in uc. In this case, the unit volume $V$ becomes the volume occupied by each sub-layer (LaO and AlO$_2$ for LaAlO$_3$ and SrO and TiO$_2$ for SrTiO$_3$), so that the unit of $n_{uc}$ is number of charge per sub-layer. Thus, $n_{eff}$, which is the total amount of charge redistribution and transfer corresponding to a particular optical transition, can be obtained by integrating $n_{uc}$ over the layer thickness, as shown in Fig. 3. (The procedure to obtain $n_{eff}$ is further explained in Methods.)

We start our discussion with insulating samples. As shown in Fig. 3a, $n_{eff}$ of A$_1$ region of LaAlO$_3$ film increases by ~0.5$e^-$, while for A$_2$ region it decreases instead, also by ~0.5$e^-$. The net amount of the charge transfer in LaAlO$_3$ film is thus (+0.5$e^-$)+(-0.5$e^-$)=0. This indicates a redistribution of ~0.5$e^-$ from O$_{Al}$-2p (AlO$_2$ sub-layer) to O$_{La}$-2p (LaO sub-layer), as shown Table 1. Based on the f-sum rule, this directly implies that there is no net charge transfer into the LaAlO$_3$/SrTiO$_3$ interface. As a result, the LaAlO$_3$/SrTiO$_3$ interface remains insulating.

Since the $LaAlO_3$ film (and the system as a whole) remains insulating, the ~0.5$e^-$ charge redistribution does not result in the creation of electrons and holes in the LaO and $AlO_2$ sub-layers but rather an increase of covalence between the LaO and $AlO_2$ sub-layers, leading to the measured charge redistribution.

One way to interpret this data is by considering that the charge redistribution is uniform for all $LaAlO_3$ layers. In this case, the covalence of $AlO_2$ becomes modified from -1 to -(1-$n_{uc}$) and the covalence of LaO from +1 to +(1-$n_{uc}$) (Fig. 4a). The $n_{uc}$ is ~0.25$e^-$–0.3$e^-$ for the 2 uc $LaAlO_3$/$SrTiO_3$, and ~0.17$e^-$–0.2$e^-$ for the 3 uc $LaAlO_3$/$SrTiO_3$ (see Methods). This charge redistribution within the $LaAlO_3$ sub-layers (electronic reconstruction[2]) can thus help to decrease the potential build-up in the $LaAlO_3$ film and partially compensate the polarization catastrophe. Combined with ionic reconstruction mechanisms such as the buckling and ionic relaxations effects predicted[22,23] and observed earlier using surface x-ray diffraction[24,25] and second harmonic generation[32], what we are measuring in terms of charge redistribution may arise from such a mechanism.

Another possible scenario that can be considered to interpret the data is that, instead of involving the whole $LaAlO_3$ layers, the charge redistribution only happens at the topmost (*i.e.*, surface) $LaAlO_3$ layer. In this case, the covalence of surface $AlO_2$ becomes modified from -1 to -0.5 and the covalence of surface LaO from +1 to +0.5, but the deeper $LaAlO_3$ layer remains unchanged, since the charge redistribution is confined only in the surface (*i.e.*, surface reconstruction). In this scenario, the surface charge redistribution is still able to partially compensate the polarization catastrophe, but it is in less of an agreement with the buckling effects, since the buckling was observed experimentally, and supported by theoretical calculations[22-25], to affect the whole $LaAlO_3$ film, not only the surface.

On the other hand, for conducting samples we observe a different phenomenon. The $n_{eff}$ of both $A_1$ and $A_2$ regions of $LaAlO_3$ film decreases (Fig. 3a). For $A_1$ region it decreases by

~0.3$e^-$–0.4$e^-$, while for A$_2$ region it decreases by ~0.2$e^-$, leading to an overall ~0.5$e^-$ decrease of $n_{eff}$ in the LaAlO$_3$ film. At the same time, for the interface layer the most significant change that happens when LaAlO$_3$/SrTiO$_3$ becomes conducting is the increase of $n_{eff}$ of B$_3$ region by ~0.5$e^-$ (Fig. 3b). The total charge transfer within the whole LaAlO$_3$/SrTiO$_3$ sample is thus again 0, with the decrease of ~0.5$e^-$ in LaAlO$_3$ film compensated by the increase of ~0.5$e^-$ at the interface. Based on the f-sum rule, this clearly indicates that there is a charge transfer of ~0.5$e^-$ from the LaAlO$_3$ film into the interface to form the 2DEG (Fig. 4b), consistent with the polarization catastrophe model[15,16]. Based on the definition of $n_{eff}$, this ~0.5$e^-$ extra charge at the interface is distributed over the ~5 nm thickness of the interface (which mostly resides in the SrTiO$_3$-side).

Moreover, it can be seen from Fig. 2c that the B$_3$ peak, which involves the transition from the new interface state that contains this ~0.5$e^-$ extra charge, is very broad (~4 eV wide), which means that the ~0.5$e^-$ is distributed over a rather wide energy range. This may be one of the reasons why transport experiments can only measure a fraction of this ~0.5$e^-$, since only a small portion of the charge is delocalized and thus able to contribute to electrical conductivity.

Furthermore, Fig. 3b also shows that the $n_{eff}$ of B$_1$ and B$_5$ regions, both of which involve transitions into the unoccupied Ti-3d-t$_{2g}$ states, decrease by ~0.05$e^-$ (*i.e.*, ~10% of 0.5$e^-$). This implies that the new interface state of conducting LaAlO$_3$/SrTiO$_3$ has Ti-3d-t$_{2g}$ characteristic[3-7,17,19,22], so the extra ~0.5$e^-$ also partially fills the previously-unoccupied Ti-3d-t$_{2g}$ state of SrTiO$_3$. This decrease is consistent with previous observations using x-ray absorption spectroscopy (XAS) experiments[3,6]. In XAS at Ti-$L_{3,2}$ edges of conducting LaAlO$_3$/SrTiO$_3$, the excitation to the unoccupied Ti-3d-t$_{2g}$ states also decreases compared to bulk SrTiO$_3$. Intriguingly, these decreases are much smaller if one assumes that all of the ~0.5$e^-$ extra charge partially fills the Ti-3d-t$_{2g}$ unoccupied DOS. This is because, based on this assumption, one would expect to observe the decrease of Ti-3d-t$_{2g}$ unoccupied DOS (and thus the $n_{eff}$ of B$_1$ and $B_5$ regions) in conducting LaAlO$_3$/SrTiO$_3$ also by an equivalent of ~0.5$e^-$. However, this is not the case, which implies that the ~0.5$e^-$ extra charge contained within the new interface state does not only reduce the number of unoccupied Ti-3d DOS, but surprisingly also other states at even higher energies, implying the importance of strong correlations and hybridizations effects in explaining the interlayer charge transfer in conducting LaAlO$_3$/SrTiO$_3$[26,42,43].

Another interesting observation to note is that in the conducting samples the $n_{\text{eff}}$ of $A_3$ region of LaAlO$_3$ film also decreases by ~0.2$e^-$. The transition in this region corresponds to O-2s state, which is strongly localized and directly corresponds to the availability of oxygen in the LaAlO$_3$ film. Thus, the decrease of O-2s DOS can indicate the presence of oxygen vacancies in the LaAlO$_3$ film of the conducting samples. For 4 uc LaAlO$_3$/SrTiO$_3$, there are 24$e^-$ in O-2s state of LaAlO$_3$, thus the ~0.2$e^-$ decrease is equivalent to ~1% oxygen vacancy. This is interesting because it has been suggested that the presence of oxygen vacancies in LaAlO$_3$ film may enhance the charge transfer from LaAlO$_3$ film into LaAlO$_3$/SrTiO$_3$ interface[7,44,45]. Because of the charge transfer into the interface, the LaAlO$_3$ film lacks ~0.5$e^-$ (*i.e.*, has additional ~0.5 holes), so the extra $e^-$ created by the oxygen vacancy may partially compensate these holes and stabilize the charge transfer. Interestingly, in insulating LaAlO$_3$/SrTiO$_3$ this oxygen vacancies signature is not observed.

It is noteworthy to reconcile our results with photoconductivity effects observed in LaAlO$_3$/SrTiO$_3$. Previous transport results[46,47] have shown that when LaAlO$_3$/SrTiO$_3$ was illuminated by photons with energies higher than the SrTiO$_3$ bandgap, its conductivity could increase due to the presence of photo-generated carriers. Based on hard XPS data[5], the amount of these photo-generated carriers is estimated to be 2.1×10$^{13}$ cm$^{-2}$ (~0.03$e^-$), which is much smaller than the number of $e^-$ contributed to the charge transfer and charge redistributions observed in our results (~0.5$e^-$). Thus, the photoconductivity effects might only influence the estimated $n_{\text{eff}}$ by ~6 %, and do not affect our analysis adversely.

Recent observations have also indicated that the cationic stoichiometry, *e.g.*, the La/Al ratio in LaAlO$_3$ film, may affect the electrical properties of LaAlO$_3$/SrTiO$_3$[18,48,49]. How this cationic stoichiometric effects would influence the high energy optical conductivity of insulating and conducting LaAlO$_3$/SrTiO$_3$ is an important open question. Thus, its interplay with the charge transfer and redistribution phenomena as observed in high energy optical conductivity still remains to be answered.

In summary, we have shown that high-energy reflectivity and spectroscopic ellipsometry studies of LaAlO$_3$/SrTiO$_3$ have revealed significant differences between the charge redistribution of insulating (2 and 3 uc of LaAlO$_3$) and charge transfer mechanisms of conducting (4 and 6 uc of LaAlO$_3$) LaAlO$_3$/SrTiO$_3$. In insulating LaAlO$_3$/SrTiO$_3$, ~0.5$e^-$ charge redistribution is observed between the AlO$_2$ and LaO sub-layers and partially compensates the polarization catastrophe. In the conducting samples, ~0.5$e^-$ is measured to be transferred from LaAlO$_3$ film into the interface, which is consistent with the polarization catastrophe model. We believe that this study reveals the nature of the intra- and inter- layer charge redistributions and charge transfers in LaAlO$_3$/SrTiO$_3$ and hence opens a path to understand the various electronic reconstructions involving the interfaces of complex oxides heterostructures. Furthermore, the use of high-energy reflectivity coupled with spectroscopic ellipsometry could be extended to other similar polar and non-polar oxide interface systems.

**Methods**

**Samples preparation.** LaAlO$_3$/SrTiO$_3$ samples were prepared by growing LaAlO$_3$ film on top of (001) SrTiO$_3$ substrates obtained from Crystec using pulsed-laser deposition (PLD)[12]. Prior to the growth, the SrTiO$_3$ substrates were treated using HF and are annealed at 950 °C for 2 hours in O$_2$ flow to achieve the desired TiO$_2$ surface termination[50]. The atomic force microscopy (AFM) topography image of the TiO$_2$ terminated SrTiO$_3$ substrate in Fig. 1a clearly shows the atomically flat surface with unit cell steps. The growth target was $LaAlO_3$ single crystal, also obtained from Crystec. The deposition pressure was $10^{-3}$ Torr, with background pressure of $10^{-9}$ Torr. The deposition temperature was 750 °C, with cooling rate of 10 °C/minute at the deposition pressure. The laser pulse frequency was 1 Hz. Four samples with varying thickness of 2, 3, 4, and 6 uc of $LaAlO_3$ film were made, as monitored using reflective high energy electron diffraction (RHEED) (Fig. 1b). After $LaAlO_3$ deposition, AFM topography measurements show that the atomically flat surface with unit cell step and terrace structure of $SrTiO_3$ is preserved, with surface roughness of ~1 Å (see Figs. 1c and 1d).

**Optics measurements.** The optical conductivity were obtained using a combination of spectroscopic ellipsometry (0.5–5.6 eV), and ultraviolet – vacuum ultraviolet (UV-VUV) reflectivity (3.7–35 eV) measurements[26-28]. The details of the optical measurements are as follow. The spectroscopic ellipsometry measurements were performed in the spectral range between 0.5 and 5.6 eV by using an SE 850 ellipsometer at room temperature[51]. Three different incident angles of 60°, 70°, and 80° from the sample normal were used, and the incident light was 45° linearly polarized from the plane of incident. For reflectivity measurements in the high-energy range between 3.7 and 35 eV we used the SUPERLUMI beamline at the DORIS storage ring of HASYLAB (DESY)[52]. The incoming photon was incident at the angle of 17.5° from the sample normal with linear polarization parallel to the sample surface. The sample chamber was outfitted with a gold mesh to measure the incident photon flux after the slit of the monochromator. The measurements were performed in ultrahigh vacuum (UHV) environment (chamber pressure of $5\times10^{-10}$ mbar) at room temperature. Prior to these measurements, the samples were heated up to 400 K in UHV to ensure that there was no additional adsorbate layers on the surface of the samples. The obtained UV-VUV reflectivity data was calibrated by comparing it with the luminescence yield of sodium salicylate ($NaC_7H_5O_3$) and the gold mesh current.

This as-measured UV-VUV reflectivity data was further normalized by using the self-normalized reflectivity extracted from spectroscopic ellipsometry[34,51], and the two normalized data were appended to obtain the combined reflectivity from 0.5–35 eV (see Supplementary Fig. 1).

**Analysis of optics data.** Both the spectroscopic ellipsometry and the combined reflectivity data was analysed using a combination of Drude-Lorentz oscillator multilayer fitting[33,34] and self-consistent iteration method (see Supplementary Methods). Due to its multilayered nature, the LaAlO$_3$/SrTiO$_3$ samples, especially the conducting cases, are considered to have three layers: the LaAlO$_3$ film on top, the SrTiO$_3$ substrate at the bottom, and the interface layer in between (see Supplementary Fig. 2), consistent with previous observation using cross-sectional conducting tip atomic force microscopy[35]. Since the spectroscopic ellipsometry data was taken at three different incident angles of 60°, 70°, and 80°, it was fitted using angle-dependent iteration method (see Supplementary Methods), and the fitting results are shown in Supplementary Figs. 3, 4, 5, 6, and 7. Supplementary Fig. 4 also shows that the thickness of the conducting interface layer is ~5 nm, consistent with previous observations[5,10,35]. Furthermore, from these variable-angle spectroscopic ellipsometry results, the absence of absorbate layer and the absence of significant anisotropy can also be inferred. On the other hand, the normalized UV-VUV reflectivity data was fitted using thickness-dependent iteration method (see Supplementary Methods), and the results of the fitting are shown in Supplementary Fig. 8.

**Complex dielectric function of LaAlO$_3$/SrTiO$_3$.** From the analysis described above, the complex dielectric function, $\varepsilon(\omega)$, of each layer of LaAlO$_3$/SrTiO$_3$ can be extracted from the high-energy reflectivity of LaAlO$_3$/SrTiO$_3$ (Fig. 2a), as presented in Fig. 5. In turn, this $\varepsilon(\omega)$ can be converted into optical conductivity $\sigma_1$ using $\sigma_1(\omega) = \varepsilon_0 \varepsilon_2(\omega) \omega$, as presented in Figs. 2b and 2c. In Figs. 5a and 5b, it can be seen that the $\varepsilon(\omega)$ of the LaAlO$_3$ film layer for both the insulating and conducting LaAlO$_3$/SrTiO$_3$ is very different than that of bulk LaAlO$_3$, which shows that the band structure of LaAlO$_3$ film is very different than bulk LaAlO$_3$. Meanwhile, the $\varepsilon(\omega)$ at the interface of the insulating samples (2 and 3 uc LaAlO$_3$/SrTiO$_3$, see Fig. 5c) is very similar to that of bulk SrTiO$_3$, which can be explained by the absence of the 2DEG in the insulating samples. Interestingly, for the conducting samples (4 and 6 uc LaAlO$_3$/SrTiO$_3$, see Fig. 5d) there are new features around 8–12 eV for $\varepsilon_1$ and 11–16 eV for $\varepsilon_2$, which, upon further analysis (see Discussions), are related to the presence of the conducting interface in those sample.

**Estimation of charge transfers and redistributions.** From equations (1) and (2), we can extract the $n_{uc}$, which is the amount of charge redistribution and transfer per uc associated with a particular optical transition relative to the bulk values. To get the accurate number, we need to carefully consider within what volume $V$ in the uc the electrons reside. Both LaAlO$_3$ and SrTiO$_3$ crystal structures can be thought of as an alternating layer structure. LaAlO$_3$ consists of alternating polar (LaO)$^+$ and (AlO$_2$)$^-$ sub-layers, while SrTiO$_3$ consists of alternating non-polar SrO and TiO$_2$ sub-layers (see Figs. 1d and 4). Due to this layered structure, in a first approximation each cation (La and Al for LaAlO$_3$, Sr and Ti for SrTiO$_3$) only occupies a volume of half uc (instead of the full one uc). For example, the La of LaAlO$_3$ has to share the space of one uc with Al (with each getting half), and likewise the Sr of SrTiO$_3$ has to share with Ti. Furthermore, for LaAlO$_3$ the valence electrons of the O atoms that belong to the two different sub-layers (LaO and AlO$_2$) contribute to two different optical transitions in the $\sigma_1$ spectra (Fig. 2b and Table 1). For simplicity, O$_{La}$ is defined as the O in the LaO plane and O$_{Al}$ as the O in the AlO$_2$ plane. Thus, O$_{La}$ also has to share the space of one uc with O$_{Al}$, with each getting the space of half uc. The same is true for SrTiO$_3$, where the O$_{Sr}$ in the SrO plane also has to share the space of one uc with the O$_{Ti}$ in the TiO$_2$ plane. This implies that the valence

17/33

electrons belonging to the different ions can also be approximated to reside in a volume of half uc. For this reason, to obtain the $n_{uc}$ the volume $V$ is chosen to be the volume of half uc of LaAlO$_3$ (lattice constant $a_0$ = 3.81 Å) or SrTiO$_3$ ($a_0$ = 3.905 Å), whichever applicable. This consideration makes the unit of $n_{uc}$ to be the number of charge per sub-layer. The result for this $n_{uc}$ estimation is shown in Fig. 6. For 2 uc LaAlO$_3$/SrTiO$_3$, the $n_{uc}$ is ~0.25$e^-$–0.3$e^-$, while for 3 uc LaAlO$_3$/SrTiO$_3$, the $n_{uc}$ is ~0.17$e^-$–0.2$e^-$. Then, the $n_{eff}$, which is the total amount of charge redistribution and transfer corresponding to a particular optical transition, can be obtained by integrating $n_{uc}$ over the layer thickness, as shown in Fig. 3.

The error bars in Figs. 3 and 6 are estimated as follows. It is assumed that there are two main sources of random errors in the data: from the resolution limitation of the optics measurements (estimated to be ~2%) and from the errors introduced in the normalization process (estimated to be ~5%). These errors affect the reflectivity data (*i.e.*, $\Delta R$), and to obtain the corresponding errors for $\sigma_1$ ($\Delta\sigma$) and thus $n_{eff}$, the errors are propagated using

$$\left|\frac{\Delta\sigma}{\sigma}\right| = \left|\frac{\Delta\varepsilon}{\varepsilon}\right| = \left|1 - \frac{1}{\sqrt{\varepsilon}} - \frac{1}{\varepsilon}\right|\left|\frac{\Delta R}{R}\right|. \tag{3}$$

**LaAlO$_3$ band structure calculation.** Unlike SrTiO$_3$ which has been studied very thoroughly[38-40], previous reports that study the band structure and high photon energy properties of LaAlO$_3$ in a detailed and comprehensive manner remain quite scarce[36,37]. Because of this, we performed our own band structure calculation of LaAlO$_3$ to complement those previous studies. The results can be used as a tool to determine the high photon energy optical transition assignments of LaAlO$_3$, as listed in Table 1.

The details of the calculation are as follows. Cubic LaAlO$_3$ has a space group of $Pm\bar{3}m$ with an experimental lattice parameters of $a = b = c = 3.8106$ Å at 821K[53]. The calculations were performed using CASTEP code[54]. Geometry optimization had been carried out with local density approximation functional (LDA) using cut-off energy of 1500 eV and a 15×15×15 Monkhorst-Pack grid[55] which corresponds to 120 **k**-points in the irreducible Brillouin zone (IBZ). The cut-off energy and **k**-point mesh had been tested and converged to energy differences of $1\times10^{-5}$ eV per atom and $4\times10^{-5}$ eV per atom, respectively. Ultrasoft pseudopotentials were generated 'on the fly' with valence states 4f, 5s, 5p, 5d, 6s for La, 3s, 3p for Al and 2s, 2p for O. The electronic minimization method used for the self-consistent field (SCF) calculation was density mixing[56] with a SCF tolerance of $2.0\times10^{-6}$ eV/atom. The geometry optimization was carried out by the Broyden–Fletcher–Goldfarb–Shanno (BFGS) algorithm[57] with energy, force and displacement tolerances of $5.0\times10^{-6}$ eV per atom, $1.0\times10^{-2}$ eV per Å, and $5.0\times10^{-4}$ Å, respectively. The optimized lattice parameter was found out to be 3.73 Å. Converged density of state calculation was carried out with a **k**-mesh of 20×20×20 and consistent with previous calculations[36,37]. The calculation results are displayed in Fig. 7.

**Acknowledgments**

We acknowledge George Sawatzky, Warren Picket, Anthony J. Leggett, Michael Coey, Daniel Khomskii, and Wei Ku for the discussions and their valuable comments. This work is supported by Singapore National Research Foundation under its Competitive Research Funding (NRF-CRP 8-2011-06 and NRF2008NRF-CRP002024), MOE-AcRF Tier-2 (MOE2010-T2-2-121), NUS-YIA, FRC, BMBF under 50KS7GUD as well as DFG through Ru 773/5-1. We acknowledge the CSE-NUS computing centre for providing facilities for our numerical calculations.

**Author Contributions**

A.R. designed the high-energy optical reflectivity coupled with spectroscopy ellipsometry experiments to study electronic reconstruction at buried interfaces. T.C.A., I.S., P.K.G., A.K., and A.R. performed high-energy optical reflectivity and spectroscopy ellipsometry measurements. A.A., A., T.V. grew high quality thin films and performed X-ray diffraction, atomic force microscopy topography and transport measurements. H.M.O. and A.R. provided the band structure calculations. T.C.A. and A.R. carried out detail data analysis and discussed the results with all co-authors. T.C.A. and A.R. wrote the paper with inputs from all co-authors. A.R. planned and supervised the project

**Additional Information**

**Competing financial interests:** The authors declare no competing financial interests.

**Correspondence** and requests for materials should be addressed to A.R. (phyandri@nus.edu.sg).

**Figures and Table**

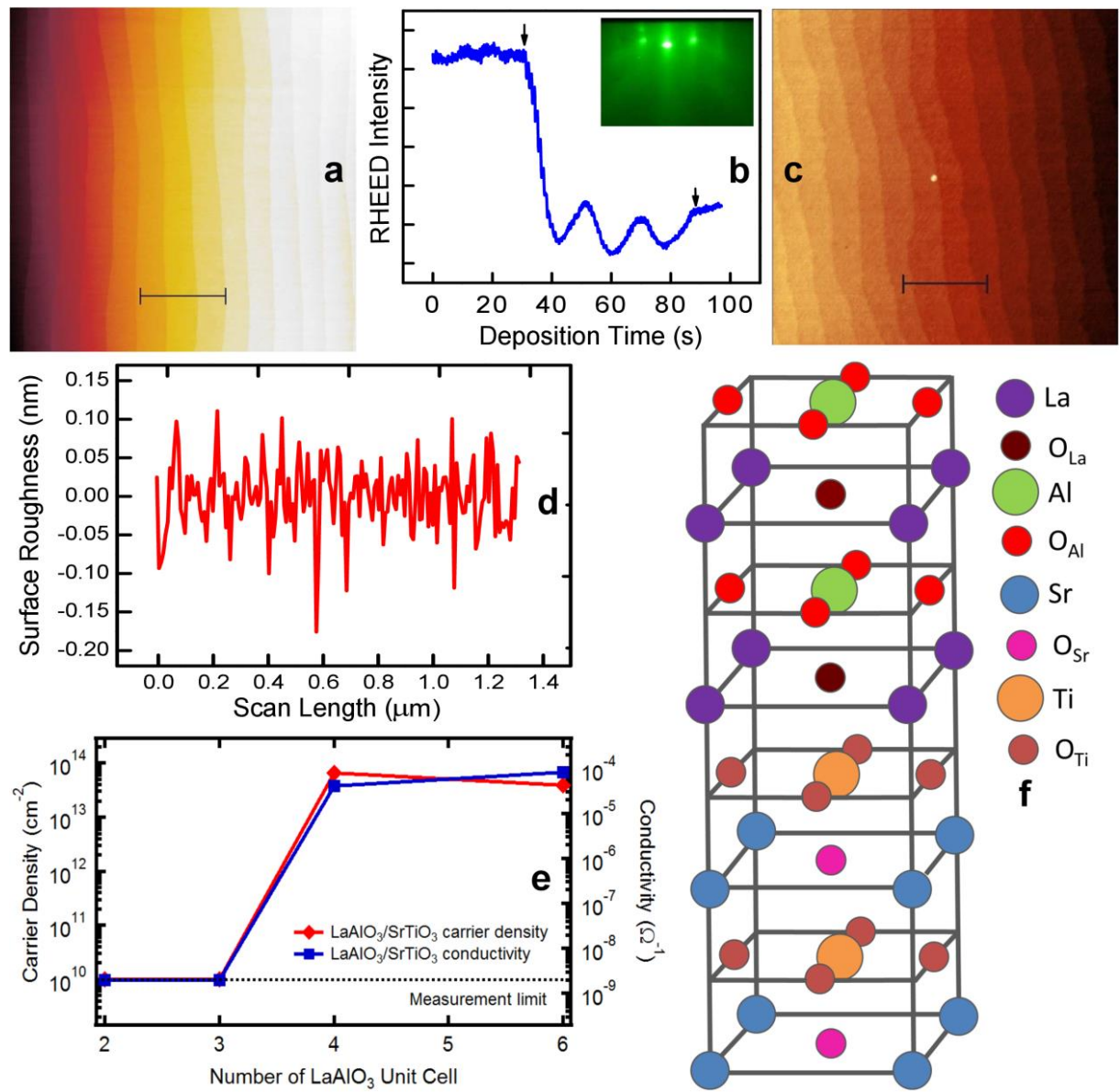

**Figure 1: Characterization results and crystal structure of LaAlO$_3$/SrTiO$_3$.** (**a**) Atomic force microscopy (AFM) topography image of TiO$_2$ terminated SrTiO$_3$ substrate. The scale bar is 1 µm. (**b**) Reflective high energy electron diffraction (RHEED) oscillations obtained for growth of 3 unit cells (uc) of LaAlO$_3$ film on SrTiO$_3$ substrate, inset shows obtained RHEED pattern after the LaAlO$_3$ growth. (**c**) AFM topography image of 4 uc LaAlO$_3$/SrTiO$_3$, showing the preserved atomically smooth surface. The scale bar is 1 µm. (**d**) The surface roughness of 4 uc LaAlO$_3$/SrTiO$_3$ as extracted from the AFM data, measured to be ~1 Å. For other LaAlO$_3$/SrTiO$_3$ samples (2, 3, and 6 uc LaAlO$_3$/SrTiO$_3$), the roughness variation and the surface AFM images are found to not alter very much as the LaAlO$_3$ thickness is below 15 uc and the layer-by-layer growth mode is preserved. (**e**) Electrical transport data of the LaAlO$_3$/SrTiO$_3$ samples as a function of LaAlO$_3$ film thickness. (**f**) Crystal structure of LaAlO$_3$/SrTiO$_3$.

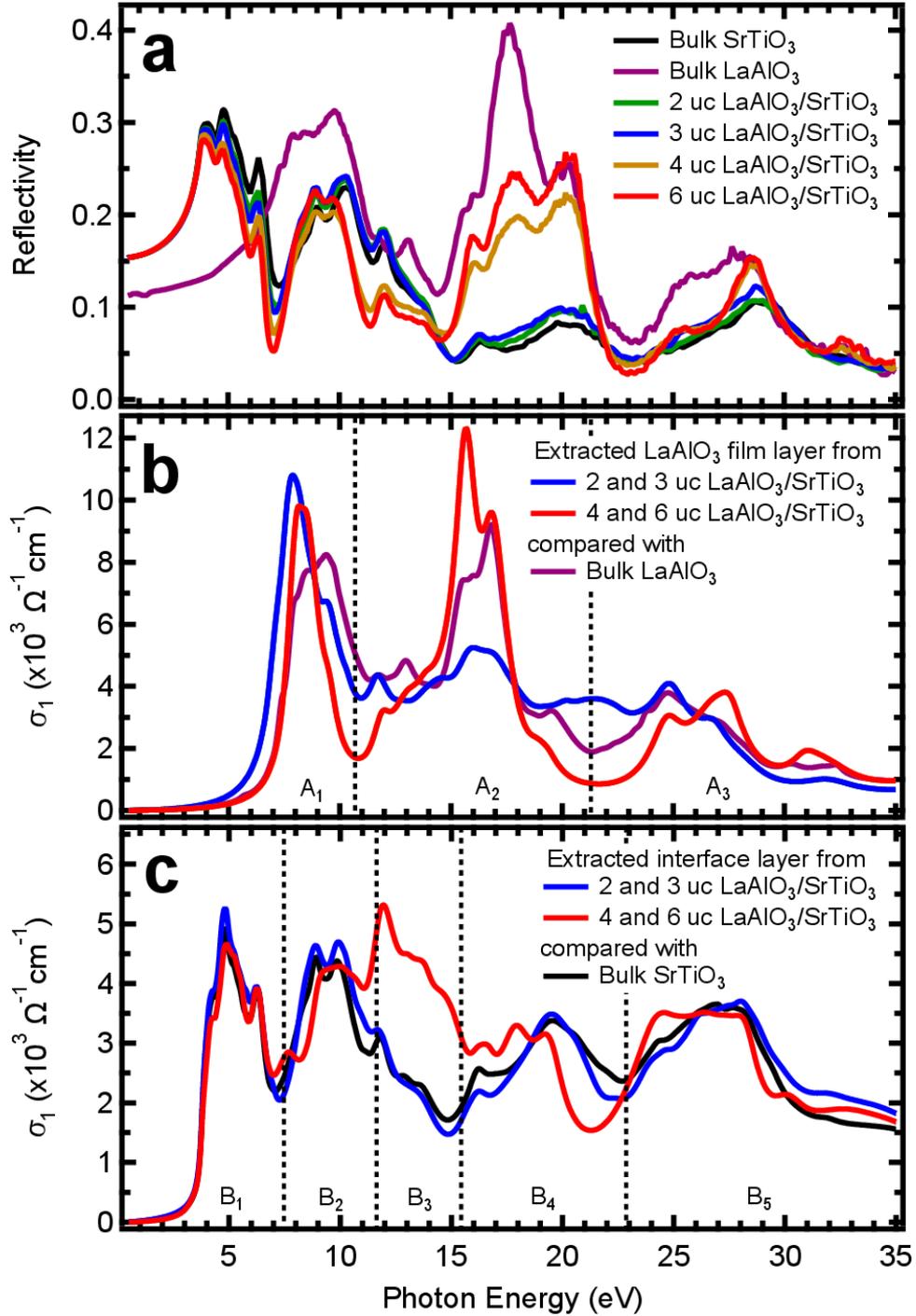

**Figure 2: Reflectivity and optical conductivity of each layer of LaAlO$_3$/SrTiO$_3$.** (**a**) Reflectivity of LaAlO$_3$/SrTiO$_3$ as compared to bulk LaAlO$_3$ and bulk SrTiO$_3$. (**b**) Extracted optical conductivity ($\sigma_1$) of LaAlO$_3$ films at different thickness of LaAlO$_3$ film, compared to bulk LaAlO$_3$. (**c**) Extracted optical conductivity of the LaAlO$_3$/SrTiO$_3$ interface at different thickness of LaAlO$_3$ film, compared to bulk SrTiO$_3$. Note that the plots for 2 and 3 unit cells (uc) LaAlO$_3$/SrTiO$_3$ are the same due to the nature of the iteration analysis used to extract $\sigma_1$ from reflectivity, and the same is true for the 4 and 6 uc LaAlO$_3$/SrTiO$_3$. The $\sigma_1$ plots are divided into several energy regions, A$_1$–A$_3$ for LaAlO$_3$ and B$_1$–B$_5$ for SrTiO$_3$ and the interface. The regions are defined based on the distinct optical transitions associated with it, which in turn based on theoretical calculations and previous reflectivity and valence electron energy loss spectroscopy[36-40].

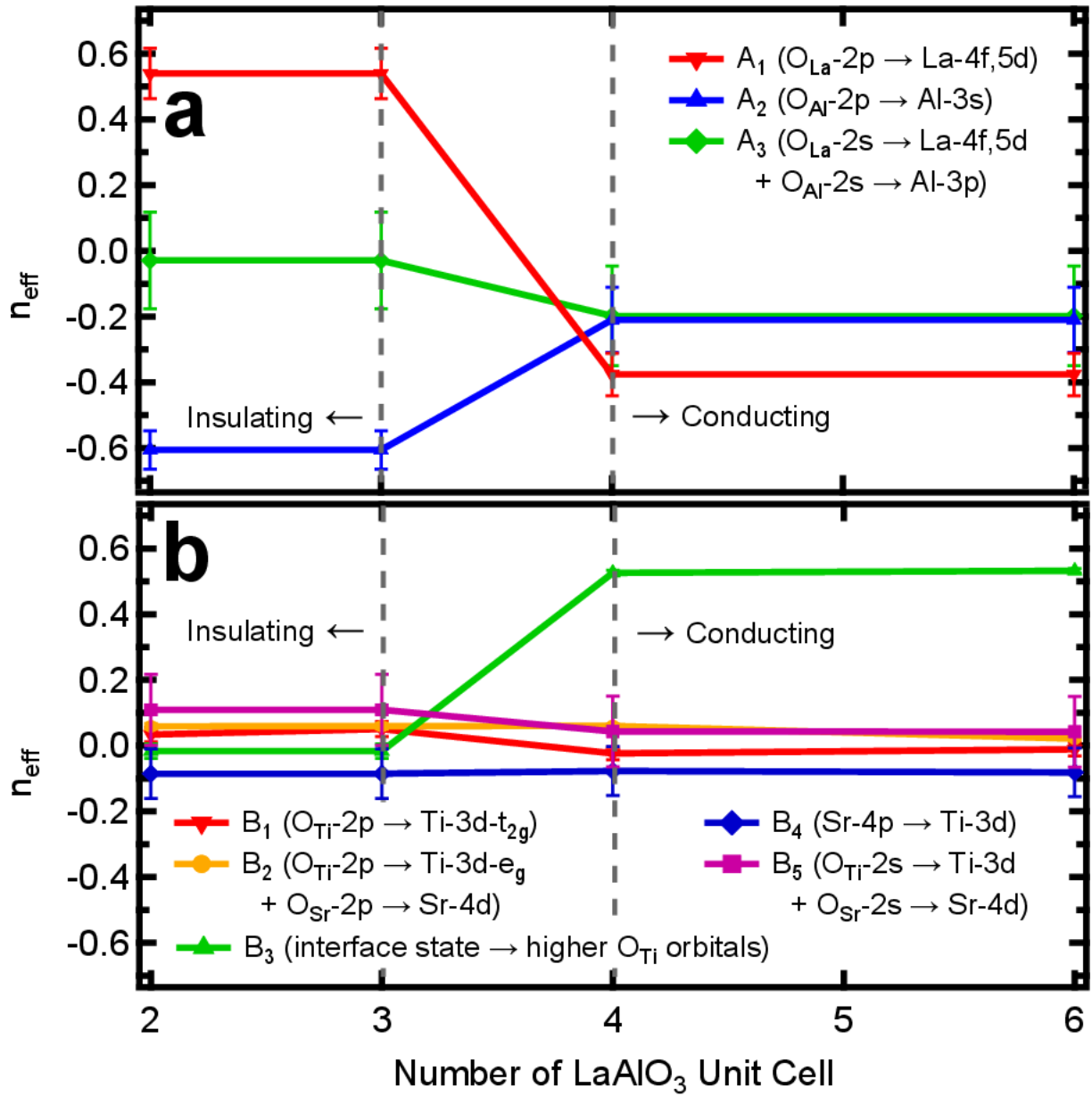

**Figure 3: The amount of charge redistribution and transfer in insulating and conducting LaAlO$_3$/SrTiO$_3$.** (**a**) The amount of charge redistribution and transfer corresponding to different energy regions in the $\sigma_1$ plots of the LaAlO$_3$ film layer, relative to bulk LaAlO$_3$ values and plotted against LaAlO$_3$ film thickness. (**b**) The amount of charge redistribution and transfer corresponding to different energy regions in the $\sigma_1$ plots of the LaAlO$_3$/SrTiO$_3$ interface layer, relative to bulk SrTiO$_3$ values and plotted against LaAlO$_3$ film thickness. Each of these energy regions can be attributed to distinct optical transitions[36-40]. The error bars are obtained from the resolution limitation of the optics measurements and the errors introduced during the reflectivity normalization procedure.

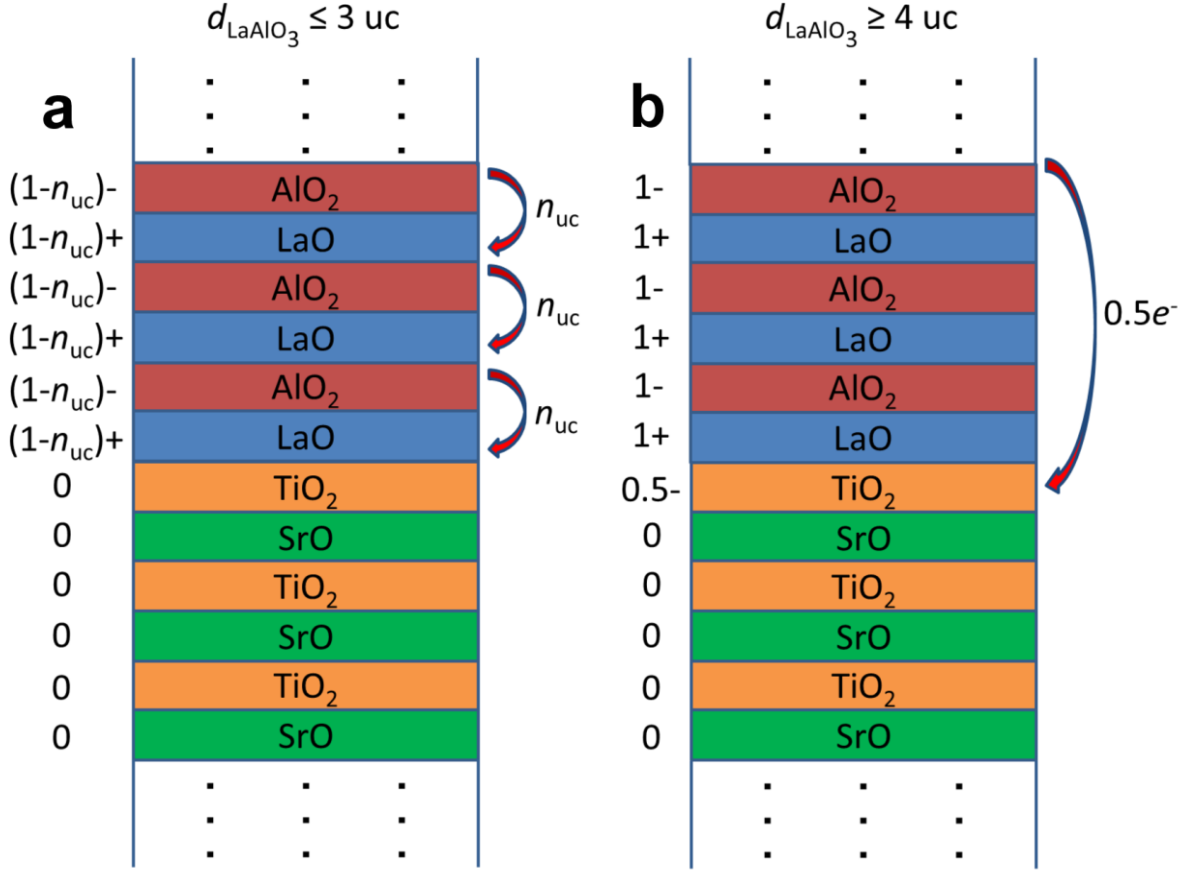

**Figure 4: Simplified pictorial layer-resolved electronic configuration model of LaAlO$_3$/SrTiO$_3$.** (**a**) Layer-resolved electronic configuration model of insulating LaAlO$_3$/SrTiO$_3$, showing the charge redistribution from AlO$_2$ sub-layer into LaO sub-layer if the redistribution is assumed to be uniform across the LaAlO$_3$ film. The $n_{uc}$ is ~0.25$e^-$–0.3$e^-$ for the 2 uc LaAlO$_3$/SrTiO$_3$, and ~0.17$e^-$–0.2$e^-$ for the 3 uc LaAlO$_3$/SrTiO$_3$. The charge redistribution can partially counteract the potential build-up due to polarization catastrophe[15,16] and keep the system insulating. (**b**) Layer-resolved electronic configuration model of conducting LaAlO$_3$/SrTiO$_3$, showing the overall charge transfer of 0.5$e^-$ from LaAlO$_3$ film into the LaAlO$_3$/SrTiO$_3$ interface, consistent with the polarization catastrophe model. In this simple picture, the extra 0.5$e^-$ is depicted to reside only within the first uc of interface, while in our results it is distributed over the ~5 nm thickness on the interface. To ensure charge conservation, one of the upper layers of LaAlO$_3$ (*i.e.*, in the dot-signed region above the third layer of LaAlO$_3$) should have AlO$_2$ sub-layer with valence state of (AlO$_2$)$^{0.5-}$ instead of (AlO$_2$)$^-$.

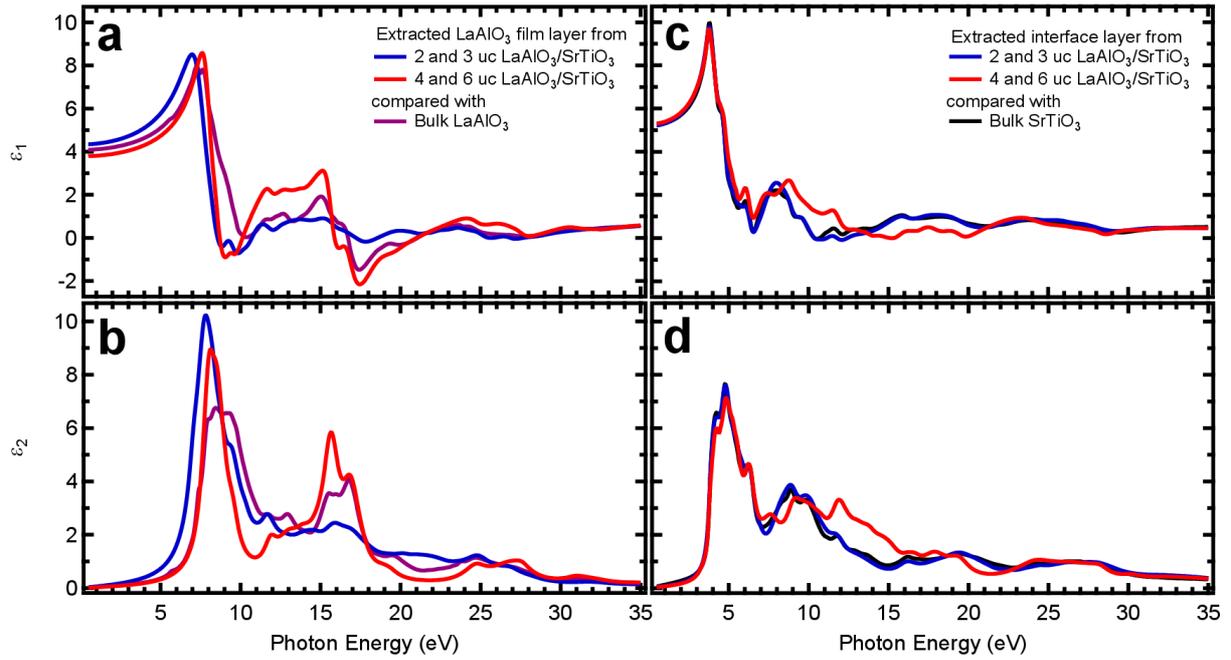

**Figure 5: Complex dielectric function of each layer of LaAlO$_3$/SrTiO$_3$.** (**a**) Real part of complex dielectric function ($\varepsilon_1$) of LaAlO$_3$ film, compared to bulk LaAlO$_3$. (**b**) Imaginary part of complex dielectric function ($\varepsilon_2$) of LaAlO$_3$ film, compared to bulk LaAlO$_3$. (**c**) The $\varepsilon_1$ of LaAlO$_3$/SrTiO$_3$ interface, compared to bulk SrTiO$_3$. (**d**) The $\varepsilon_2$ of LaAlO$_3$/SrTiO$_3$ interface, compared to bulk SrTiO$_3$. Note that the plots for 2 and 3 unit cells are the same due to the nature of the thickness-dependent iteration, and the same is true for the 4 and 6 unit cells case.

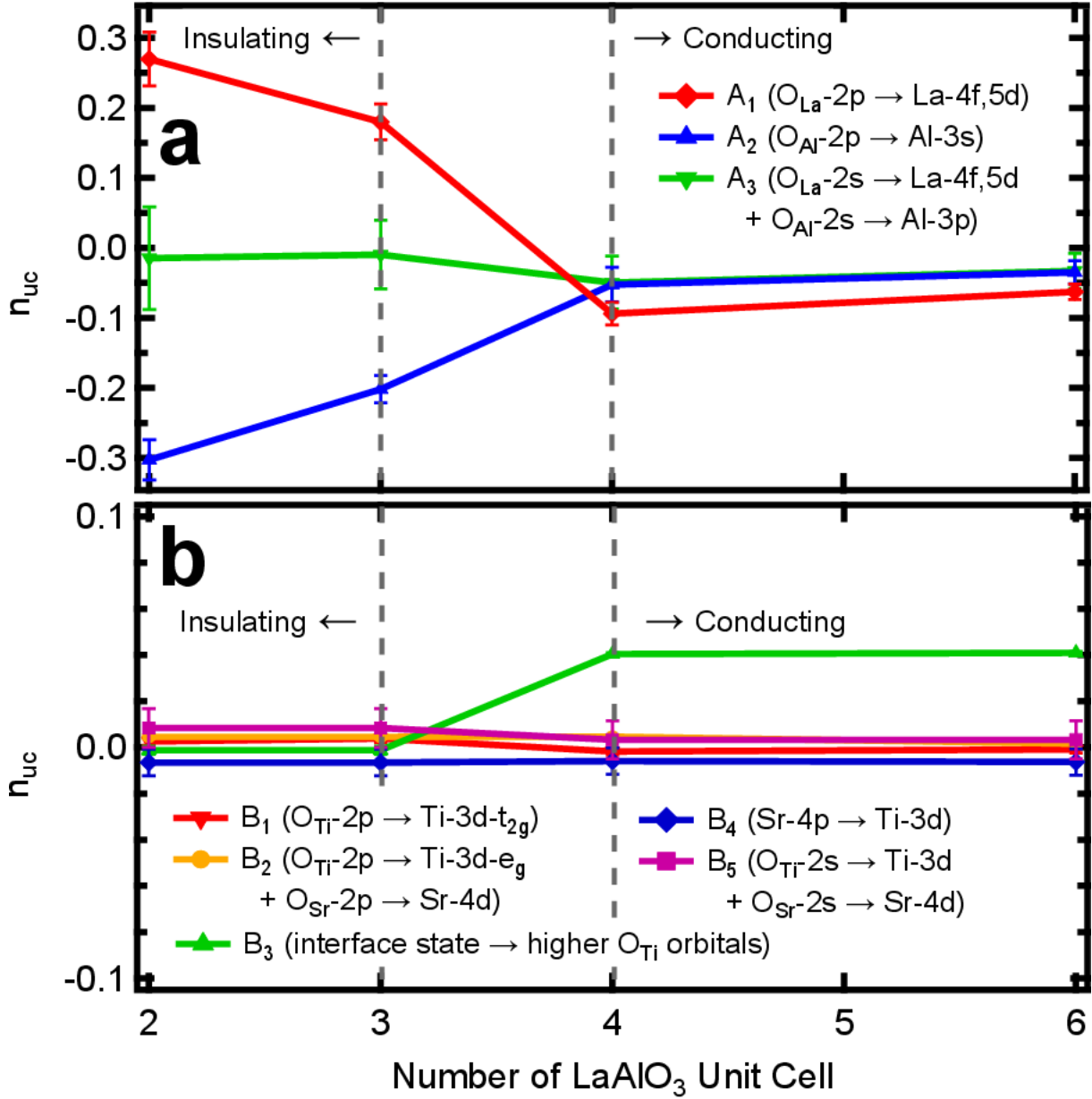

**Figure 6: Effective number of charge per unit cell of LaAlO$_3$ and interface layer of LaAlO$_3$/SrTiO$_3$.** (**a**) The effective number of charge per unit cell, $n_{uc}$, of LaAlO$_3$ film, if the charge distribution is assumed to be uniform over the LaAlO$_3$ thicknesses (2, 3, 4, and 6 unit cells). (**b**) The $n_{uc}$ of the LaAlO$_3$/SrTiO$_3$ interface, if the charge distribution is assumed to be uniform over the interface thickness (~5.3 nm). Each of these energy regions can be attributed to distinct optical transitions[36-40]. The error bars are obtained from the resolution limitation of the optics measurements and the errors introduced during the reflectivity normalization procedure.

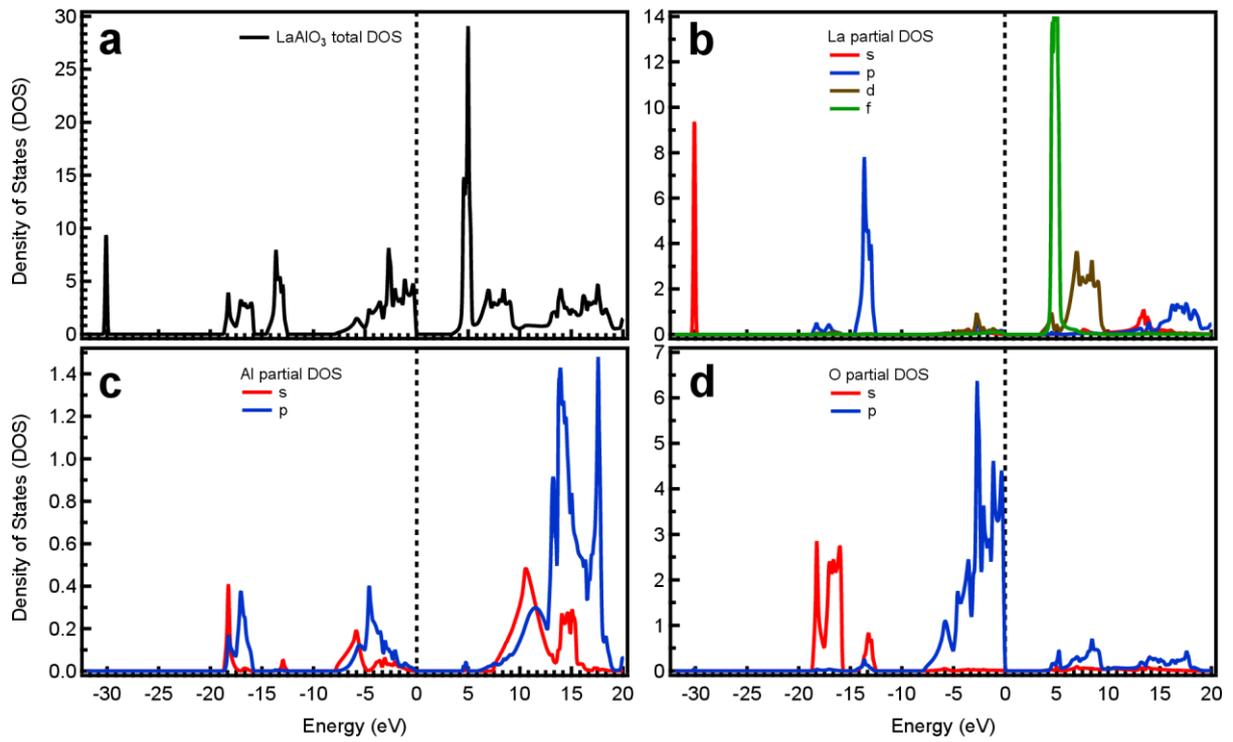

**Figure 7: Theoretical band structure of LaAlO₃.** (**a**) Total density of states (DOS) of LaAlO$_3$. (**b**) Partial DOS of La. (**c**) Partial DOS Al. (**d**) Partial DOS of O. The dotted lines are the Fermi level.

**Table 1: Main optical transitions of bulk LaAlO₃ and bulk SrTiO₃.** The assignments are based on theoretical calculations and previous reflectivity and valence electron energy loss spectroscopy[36-40]. Note that the transition $B_3$ does not exist in bulk SrTiO$_3$; rather it is a new transition that arises from the new interface state at the conducting interface of LaAlO$_3$/SrTiO$_3$ as a characteristic of the two-dimensional electron gas (2DEG), which includes the newly-occupied Ti-3d-$t_{2g}$ states.

| Region | Main Optical Transition | Photon Energy (eV) |
|---|---|---|
| LaAlO$_3$ | | |
| $A_1$ | O$_{La}$-2p → La-4f,5d | 0.5 – 10.6 |
| $A_2$ | O$_{Al}$-2p → Al-3s | 10.6 – 21.5 |
| $A_3$ | O$_{La}$-2s → La-4f,5d & O$_{Al}$-2s → Al-3p | 21.5 – 35.0 |
| SrTiO$_3$ | | |
| $B_1$ | O$_{Ti}$-2p → Ti-3d-$t_{2g}$ | 0.5 – 7.1 |
| $B_2$ | O$_{Ti}$-2p → Ti-3d-$e_g$ & O$_{Sr}$-2p → Sr-4d | 7.1 – 11.3 |
| $B_3$ (only occurs at conducting interface) | interface state → higher O$_{Ti}$ orbitals | 11.3 – 15.1 |
| $B_4$ | Sr-4p → Ti-3d | 15.1 – 22.7 |
| $B_5$ | O$_{Ti}$-2s → Ti-3d & O$_{Sr}$ → Sr-4d | 22.7 – 35.0 |

## Supplementary Materials

**Supplementary Methods**

**General analysis of optics data of bulk materials.** We first discuss the general analysis of the spectroscopic ellipsometry data (0.5–5.6 eV) of bulk materials. The spectroscopy ellipsometry is a self-normalizing technique to determine the complex element of dielectric tensor from a single measurement without performing a Kramers-Kronig transformation, making it free from any ambiguities that are related to the normalization of conventional reflectivity results[1]. The raw data measured by ellipsometry is expressed in terms of $\Psi$ (change in intensity) and $\Delta$ (change in phase), which are defined as[2]

$$\tan \Psi \exp(i\Delta) \equiv \frac{r_p}{r_s}. \tag{1}$$

where $r_p$ and $r_s$ are the reflectivity of p- (parallel to the plane of incident) and s- (perpendicular to the plane of incident) polarized light. From the Fresnel equation, these two quantities can be defined as

$$r_p^{ij} = \frac{n_j \cos\theta_i - n_i \cos\theta_j}{n_j \cos\theta_i + n_i \cos\theta_j} \tag{2}$$

and

$$r_s^{ij} = \frac{n_i \cos\theta_i - n_j \cos\theta_j}{n_i \cos\theta_i + n_j \cos\theta_j} \tag{3}$$

Here, $n$ and $\theta$ represent the refraction index and angle of incident from the surface normal, respectively. The $i$ and $j$ represent the two materials involved in the photon propagation. From here, the complex dielectric function $\varepsilon(\omega)$ can be obtained using

$$\sqrt{\varepsilon(\omega)} = n(\omega), \qquad (4)$$

where $\omega$ is the photon frequency. The $\varepsilon(\omega)$ obtained using supplementary equation (S4) can then be converted back to reflectivity using supplementary equations (2) and (3). The tan $\Psi$ and $\Delta$ are essentially *ratios* of the intensities (for $\Psi$) and phases (for $\Delta$) of the reflectivity of the p- and s-polarized lights (supplementary equation (1)), which makes them (and any quantities derived from them, including reflectivity) *self-normalized*.

Next, we discuss the analysis of the UV-VUV reflectivity data (3.7–35 eV) of bulk materials. Due to the self-normalized nature of spectroscopic ellipsometry, the ellipsometry-derived reflectivity can be used to normalize the UV-VUV reflectivity at the low energy side within the range of 3.7–5.6 eV. (In this case, because the light polarization used in UV-VUV reflectivity experiment is parallel to the sample surface, which is equivalent to perpendicular to the plane of incident, supplementary equation (3) is used to calculate the reflectivity in the ellipsometry range.) Furthermore, the high energy part (>30 eV) is normalized using calculations based on off-resonance scattering considerations according to[3]

$$r = i \frac{r_0 \lambda}{\sin \theta} F(\theta) P_f(2\theta), \qquad (5)$$

where $r$ is the reflectivity, $r_0$ is the classical electron radius ($\frac{e^2}{mc^2}$), $\lambda$ is the photon wavelength, $P_f(\theta)$ is the polarization factor (equal to unity for s-polarized light and equal to $\cos \theta$ for p-polarized light), and $F(\theta)$ is the structure factor per unit area given by

$$F(\theta) = \sum_q n_q f_q \exp\left(\frac{i 4\pi z_q}{\lambda} \sin \theta\right). \qquad (6)$$

The summation is performed over the different types of atoms on a particular atomic plane on which the light is incident, with $n_q$ denotes the number of atoms of type $q$ in that particular plane, $f_q$ denotes the tabulated atomic form factor corresponding to that atom $q$, and $z_q$ denotes the direction vector normal to the plane in question. From the above step, normalized reflectivity in the range of 0.5–35 eV can be obtained. As an example for the normalization procedure, Supplementary Fig. 1 shows the normalized high-energy reflectivity (0.5–35 eV) of SrTiO$_3$ as compared to the self-normalized reflectivity obtained from spectroscopic ellipsometry (0.5–5.6 eV), the unnormalized UV-VUV reflectivity (3.7–35 eV), and the off-resonance considerations (>30 eV).

For nearly isotropic materials like bulk LaAlO$_3$ and bulk SrTiO$_3$, the dielectric function $\varepsilon(\omega) = \varepsilon_1(\omega) + i\varepsilon_2(\omega)$ can be extracted from the normalized UV-VUV reflectivity using Kramers-Kronig analysis[4,5], which is based on the Kramers-Kronig relationship between the real and imaginary parts of $\varepsilon(\omega)$,

$$\varepsilon_1(\omega) - 1 = -\frac{2}{\pi} P \int_0^\infty \frac{x\varepsilon_1(\omega)}{\omega^2 - x^2} dx \tag{7}$$

and

$$\varepsilon_2(\omega) = -\frac{2\omega}{\pi} P \int_0^\infty \frac{\varepsilon_1(\omega) - 1}{\omega^2 - x^2} dx, \tag{8}$$

where $P$ is the Cauchy principal value. In our case, the analysis is done by fitting using the Kramers-Kronig-transformable Drude-Lorentz oscillators according to

$$\varepsilon(\omega) = \varepsilon_\infty + \sum_k \frac{\omega_{p,k}}{\omega_{0,k}^2 - \omega^2 - i\Gamma_k \omega}. \tag{9}$$

The high frequency dielectric constant is denoted by $\varepsilon_\infty$; $\omega_{p,k}$, $\omega_{0,k}$, and $\Gamma_k$ are the plasma frequency, the transverse frequency (eigen frequency), and the line width (scattering rate) of the $k$-th oscillator, respectively. Since the energy range involved is very broad (covering 0.5–35 eV), the analysis yields a stabilized Kramers-Kronig transformation.

**Multilayer analysis of spectroscopic ellipsometry data assisted with self-consistent iteration.** LaAlO$_3$/SrTiO$_3$ heterostructure is layered along the <001> direction (perpendicular to the

(001) surface of the sample) due to its heterostructure nature as well as the presence of the conducting layer at $LaAlO_3/SrTiO_3$ interface. For this reason, a multilayer consideration based on a boundary analysis of Fresnel equation needs to be taken into account in analyzing the spectroscopic ellipsometry data (and later also the UV-VUV reflectivity) of $LaAlO_3/SrTiO_3$. In this multilayer analysis, the conducting $LaAlO_3/SrTiO_3$ has three layers: $LaAlO_3$ film layer on top, bulk $SrTiO_3$ substrate at the bottom, and an interface layer sandwiched in between, representing the 2DEG of the conducting samples, consistent with previous observation using cross-sectional conducting tip atomic force microscopy[6] (Supplementary Fig. 2). For the sake of consistency, in the initial model of insulating $LaAlO_3/SrTiO_3$ the interface layer is retained, and later on after the iteration converges we obtain that the interface has optical properties similar to bulk $SrTiO_3$. This makes the insulating $LaAlO_3/SrTiO_3$ an effective two-layer structure instead, due to the absence of the conducting interface layer. It should be noted that in this layered structure, the 2DEG is assumed to have uniform distribution over a certain interface layer thickness. According to previous reports[7-9], the apparent distribution of the 2DEG is more likely to be asymmetric: the charge density is highest right at the interface, then it slowly decays as it goes deeper into the $SrTiO_3$ substrate. Furthermore, due to the strong temperature dependence of dielectric constant of $SrTiO_3$[10], the interface layer thickness may likewise also have some temperature dependence.

According to analysis of wave propagation in a stratified medium, the reflectivity of a three-layer system can be expressed through Fresnel equations as[2,11]

$$r_{vac,multi} = \frac{r_{vac,LaAlO_3} + r_{LaAlO_3,int}e^{i2\delta_{LaAlO_3}} + r_{vac,LaAlO_3}r_{LaAlO_3,int}r_{int,SrTiO_3}e^{i2\delta_{int}} + r_{int,SrTiO_3}e^{i2(\delta_{LaAlO_3}+\delta_{int})}}{1 + r_{vac,LaAlO_3}r_{LaAlO_3,int}e^{i2\delta_{LaAlO_3}} + r_{LaAlO_3,int}r_{int,SrTiO_3}e^{i2\delta_{int}} + r_{vac,LaAlO_3}r_{int,SrTiO_3}e^{i2(\delta_{LaAlO_3}+\delta_{int})}}$$

,(S10)

where

$$\delta_i = \frac{2\pi d_i}{\lambda} n_i \cos\theta_i. \tag{S11}$$

Here, $d$ represents the layer thickness (of both the LaAlO$_3$ film and interface layer), while the subscripts vac, multi, and int represent the vacuum, the LaAlO$_3$/SrTiO$_3$ multilayer, and the interface layer, respectively, which, along with LaAlO$_3$ and SrTiO$_3$, are the various materials involved in the propagation of the photon.

As seen from supplementary equation (10), the reflectivity (and, by extension via supplementary equation (1), Ψ and Δ) of LaAlO$_3$/SrTiO$_3$ contains mixed information from all the three constituent layers. This makes the extraction of the $\varepsilon(\omega)$ of individual layer non-trivial, because it cannot be converted directly from Ψ and Δ like bulk materials, due to the too many unknown factors involved. Since $\varepsilon(\omega)$ of the bulk SrTiO$_3$ substrate can be measured independently (Fig. 2b) and LaAlO$_3$ film thicknesses ($d_{\text{LaAlO3}}$) are known from RHEED measurements during sample growths, supplementary equations (1) – (4), (10), and (11) left us with 3 unknown variables: $\varepsilon(\omega)$ of LaAlO$_3$ film (which might be different from that of bulk LaAlO$_3$), $\varepsilon(\omega)$ of interface layer, and the thickness of the interface layer ($d_{\text{int}}$). (Note that even though $\varepsilon(\omega)$ has real and imaginary parts, they are connected through the Kramer-Kronig relationships, see supplementary equations (7) and (8)). Solving this is mathematically non-trivial, since (assuming there is no change in $\varepsilon(\omega)$ of bulk SrTiO$_3$ across the samples) there are 3 unknowns but only 1 equation (supplementary equation (1)). To overcome this, supplementary equation (1) can be diversified by using supplementary equation (11), for example by varying the angle of incident $\theta$. In spectroscopic ellipsometry (0.5–5.6 eV), this can be done by measuring Ψ and Δ at 3 different incident angles: 60°, 70°, and 80° from the sample normal. This results in 3 sets of Ψ and Δ data, which provides us the 3 equations necessary to perform a self-consistent iteration procedure to extract the 3 unknown variables.

As the representative, for the 4 uc LaAlO$_3$/SrTiO$_3$, the iteration for the spectroscopic ellipsometry data can be performed as following. Previous studies report that the thickness of the conducting interface might be around 2–7 nm[6-9,12-14], so the initial guess for $d_{int}$ can be reasonably set as 5 nm. Meanwhile, the initial guess for $\varepsilon(\omega)$ of LaAlO$_3$ film can be set as the same as $\varepsilon(\omega)$ of bulk LaAlO$_3$, which can be obtained independently (Fig. 2b). With these two variables fixed, supplementary equation (1) is fitted into experimental value of Ψ and Δ measured at $\theta$=60° using supplementary equation (9) by appropriately adjusting the Drude-Lorentz oscillators that make up the $\varepsilon(\omega)$ of interface layer. Then, the newly-fitted $\varepsilon(\omega)$ of interface is fixed, and $d_{int}$ is appropriately adjusted so that supplementary equation (1) can now be fitted into experimental value of Ψ and Δ measured at $\theta$=70°. After that, the newly-adjusted $d_{int}$ is also fixed (along with the previously-fitted $\varepsilon(\omega)$ of interface), and supplementary equation (1) is fitted into experimental value of Ψ and Δ measured at $\theta$=80° using supplementary equation (9) by appropriately adjusting the Drude-Lorentz oscillators that make up the $\varepsilon(\omega)$ of LaAlO$_3$ film layer. The process is then repeated by going back to Ψ and Δ values measured at $\theta$=60° and subsequently cycling through the incident angles, fitting only 1 variable each step while keeping the other 2 fixed. Convergence is reached when $\varepsilon(\omega)$ of LaAlO$_3$ film, $\varepsilon(\omega)$ of interface layer, and $d_{int}$ can satisfy supplementary equation (1) for all three incident angles, as shown in Supplementary Fig. 3. In other words, the iteration results form a universal fitting that can match the data from all incident angles. It can be clearly seen that without the interface layer, the fitted Ψ and fitted Δ do not match the measured Ψ and measured Δ, *i.e.*, the universal fitting cannot be achieved. The iteration thus results in the converged values of thickness of interface layer, $\varepsilon(\omega)$ of LaAlO$_3$ film, and $\varepsilon(\omega)$ of interface layer, as shown in Fig. 5. Along with the already-known $\varepsilon(\omega)$ of bulk SrTiO$_3$ and thickness of LaAlO$_3$ film, these quantities can be converted to reflectivity in the 0.5–5.6 eV range using supplementary equation (3), which then can be used to normalize the UV-VUV reflectivity.

Supplementary Fig. 4a illustrates the iteration process by showing the evolution of $d_{int}$ through each iteration step. As it can be seen, as the iteration progresses the value of $d_{int}$ slowly approaches a distinct asymptotic value, and at step 5 it finally converges into ~5.2 nm.

For 6 uc LaAlO$_3$/SrTiO$_3$, the iteration process can be performed similarly, since the only difference is $d_{LaAlO3}$, which is known and can be appropriately adjusted using supplementary equation (11). Supplementary Fig. 5 show the fitted $\Psi$ and fitted $\Delta$ after convergence that match the measured $\Psi$ and measured $\Delta$. Again, without the interface layer, the universal fitting cannot be achieved. The iteration progress of $d_{int}$ for 6 uc LaAlO$_3$/SrTiO$_3$ is shown in Supplementary Fig. 4b. In this case, the initial guess for $d_{int}$ is set to be 6 nm, and the final converged value is found to be ~5.3 nm, very close to the 4 uc value of ~5.2 nm. This indicates that the properties of 4 and 6 uc LaAlO$_3$/SrTiO$_3$ are very similar, and any apparent differences in $\Psi$, $\Delta$, and reflectivity values between the two samples are mainly caused by the difference in $d_{LaAlO3}$. In fact, because of this, since from 4 uc iteration the converged values for $\varepsilon(\omega)$ of LaAlO$_3$ film, $\varepsilon(\omega)$ of interface layer, and $d_{int}$ of 4 uc LaAlO$_3$/SrTiO$_3$ are already obtained, those values can also be used as the initial guess for the 6 uc LaAlO$_3$/SrTiO$_3$ iteration. It can be seen from Supplementary Fig. 4b that with those better starting points, the iteration process can be simplified and convergence can be achieved with fewer steps, while still reaching the same converged value of $d_{int}$=~5.3 nm. This confirms the self-consistency of the iteration process, showing that even if it starts with different initial guesses, the iteration does eventually converge into the same final results. The converged values from the iteration can then be converted into reflectivity using supplementary equation (3), which then will be used to normalize the UV-VUV reflectivity.

For the insulating 2 uc LaAlO$_3$/SrTiO$_3$, the iteration-based analysis is also performed similarly. The fitted $\Psi$ and fitted $\Delta$ after convergence that match the measured $\Psi$ and measured $\Delta$ are shown in Supplementary Fig. 6. For the sake of consistency and to make layer-by-layer comparison between insulating and conducting LaAlO$_3$/SrTiO$_3$ more readily apparent, the interface layer is still initially retained in the iteration process. However, as shown later, after the multilayer analysis of the combined ellipsometry-derived and UV-VUV reflectivity, the $\varepsilon(\omega)$ of the (artificial) interface layer is found to be very similar to that of bulk SrTiO$_3$, making insulating LaAlO$_3$/SrTiO$_3$ effectively a two-layer structure instead. This can be explained by the absence of the conducting interface in insulating LaAlO$_3$/SrTiO$_3$. Similar with the conducting samples, the resulting $\varepsilon(\omega)$ can be converted into reflectivity using supplementary equation (3) to be used in UV-VUV reflectivity normalization process. The same is true for the insulating 3 uc LaAlO$_3$/SrTiO$_3$ (see Supplementary Fig. 7).

**Multilayer analysis of reflectivity (from 0.5 to 35 eV) assisted with self-consistent iteration.** From the iteration-based analysis of the spectroscopic ellipsometry data, the $\varepsilon(\omega)$ of each individual constituent layer of LaAlO$_3$/SrTiO$_3$, along with their thicknesses, can be extracted. These quantities can then be converted into the reflectivity in the 0.5–5.6 eV range using supplementary equation (3). From here, the normalization of the UV-VUV reflectivity data (3.7–35 eV) is performed similar to bulk materials: using the ellipsometry-derived reflectivity to normalize the low energy side (3.7–5.6 eV) and the off-resonance scattering considerations according to supplementary equations (5) and (6) to normalize the high energy side (>30 eV). Then, the normalized reflectivity can be used to extract the $\varepsilon(\omega)$ of the constituent materials.

A similar procedure in the analysis of spectroscopic ellipsometry data of LaAlO$_3$/SrTiO$_3$ applies also in analyzing the high-energy reflectivity data, but this time we vary the LaAlO$_3$ thicknesses. Even though $d_{\text{int}}$ is already known to be ~5.3 nm from the spectroscopic ellipsometry iteration analysis, it still leaves us with 2 unknowns (high photon energy $\varepsilon(\omega)$ of LaAlO$_3$ film and high photon energy $\varepsilon(\omega)$ of interface layer) but only 1 equation (sup-plementary equation (10)), which prevents a straight-forward mathematical solution. Furthermore, due to a fixed incident angle of 17.5° from the sample normal, similar angle-dependent iteration as the one done in the spectroscopic ellipsometry region cannot be performed. To address this, we note that supplementary equation (10) can also be diversified through supplementary equation (11) by varying the layer thickness, in particular the $LaAlO_3$ film thickness $d_{LaAlO3}$. It is for this reason that we have intentionally fabricated a pair of insulating samples (2 and 3 unit cells of $LaAlO_3$) and a pair of conducting samples (4 and 6 unit cells of $LaAlO_3$). Each pair has similar respective physics, with only difference in $d_{LaAlO3}$, which can be taken care of by appropriately adjusting supplementary equation (11). This means for each case (insulating and conducting), there are 2 unknowns and 2 equations for $r_{vac,multi}$, so a self-consistent iteration can be used to extract $\varepsilon(\omega)$ of each individual layer.

The iteration process is essentially similar to the angle-dependent iteration, only instead of iterating through different data obtained using different incident angles, it is done by iterating through reflectivity data obtained from samples with different $LaAlO_3$ film thicknesses. Take the conducting samples for example. The iteration is done by alternating between the reflectivity data of 4 and 6 uc samples, only fitting one layer (using supplementary equation (9)) at a time while keeping the other constant. Convergence is reached when $\varepsilon(\omega)$ of $LaAlO_3$ film and $\varepsilon(\omega)$ of interface layer can satisfy supplementary equation (10) for both the 4 and 6 uc samples. The insulating samples are also analyzed using the same procedure. Again, for the sake of consistency and to make layer-by-layer comparison between insulating and conducting $LaAlO_3/SrTiO_3$ more readily apparent, the interface layer is also initially retained in the iteration process of the insulating samples. After convergence, the $\varepsilon(\omega)$ of the (artificial) interface layer of the insulating samples is again found to be very similar to that of bulk $SrTiO_3$ due to the absence of the conducting interface, making insulating $LaAlO_3/SrTiO_3$ effectively a two-layer structure instead. From Supplementary Fig. 8, shows the results of the analysis, showing the fitted reflectivity of each sample as compared to the measured values, and it can be seen that generally the fitted reflectivity of each sample is matched very closely with its respective measured reflectivity.

An important requirement for this thickness-dependent iteration method is that any variation in reflectivity among the samples should be only due to the different film thicknesses involved (*i.e.*, only due to supplementary equation (11)). Film thickness differences should not significantly modified the internal properties of the samples, because otherwise it will make the iteration procedure impossible. For the LaAlO$_3$/SrTiO$_3$, the thickness-dependent iteration is performed between the 2 and 3 uc LaAlO$_3$/SrTiO$_3$ samples to resolve the $\varepsilon(\omega)$ of LaAlO$_3$ film and the interface layer in the insulating case, and between the 4 and 6 uc LaAlO$_3$/SrTiO$_3$ samples to resolve the $\varepsilon(\omega)$ of LaAlO$_3$ film and the interface layer in the conducting case. These treatments are based on the observations that the reflectivity of 2 and 3 uc LaAlO$_3$/SrTiO$_3$ is similar to each other (see Fig. 2a), and by simply varying the LaAlO$_3$ film thickness appropriately, the reflectivity of 2 and 3 uc LaAlO$_3$/SrTiO$_3$ can be fitted with one universal set of Drude-Lorentz parameters. Likewise is true for 4 and 6 uc LaAlO$_3$/SrTiO$_3$.

The requirement is also why the iteration cannot be performed between the insulating 3 uc LaAlO$_3$/SrTiO$_3$ and the conducting 4 uc LaAlO$_3$/SrTiO$_3$, since their inherent properties (such as the conductivity of the interface) are modified by the increase of the LaAlO$_3$ thickness. Furthermore, due to this requirement and as an inherent consequence of the thickness-dependent iteration, the resulting $\varepsilon(\omega)$ (and thus optical conductivity $\sigma_1(\omega) = \varepsilon_0 \varepsilon_2(\omega) \omega$) of each layer is identical between the samples that are being iterated. For instance, $\sigma_1$ of the LaAlO$_3$ film shown in Fig. 2b is the same for both the 2 and 3 uc LaAlO$_3$/SrTiO$_3$, and the two cannot be distinguished. The same is also true for the interface layer of 2 and 3 uc LaAlO$_3$/SrTiO$_3$ (Fig. 2c), and for both layers of 4 and 6 uc LaAlO$_3$/SrTiO$_3$.

**Supplementary References**

S12. Reyren, N. *et al*. Superconducting interfaces between insulating oxides. *Science* **317**, 1196–1199 (2007).

S13. Siemons, W. *et al*. Origin of charge density at LaAlO$_3$ on SrTiO$_3$ heterointerfaces: Possibility of intrinsic doping. *Phys. Rev. Lett.* **98**, 196802 (2007).

S14. Sing, M. *et al.* Profiling the interface electron gas of LaAlO$_3$/SrTiO$_3$ heterostructures with hard X-ray photoelectron spectroscopy. *Phys. Rev. Lett*. **102**, 176805 (2009).

**Supplementary Figures**

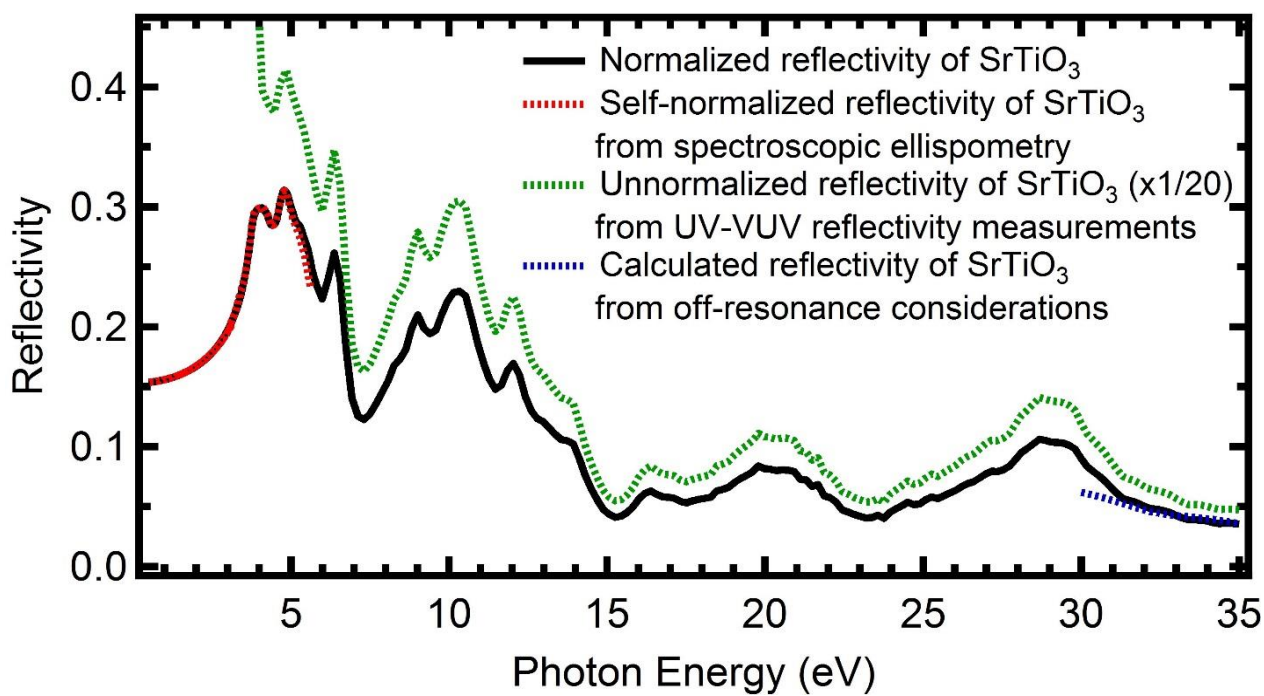

**Supplementary Figure 1: Comparison between normalized and unnormalized reflectivity of bulk SrTiO$_3$.** The normalized high-energy reflectivity (0.5–35 eV) of SrTiO$_3$ is compared to the self-normalized reflectivity obtained from spectroscopic ellipsometry (0.5–5.6 eV), the unnormalized UV-VUV reflectivity (3.7–35 eV) from the UV-VUV reflectivity measurements (scaled down by 20x to fit the graph), and the calculated reflectivity from off-resonance considerations (>30 eV). The unnormalized reflectivity is normalized by further scaling it down to match the spectroscopic ellipsometry and the off-resonance data, and then the three data are appended together to obtained the normalized reflectivity in the full range of 0.5–35 eV.

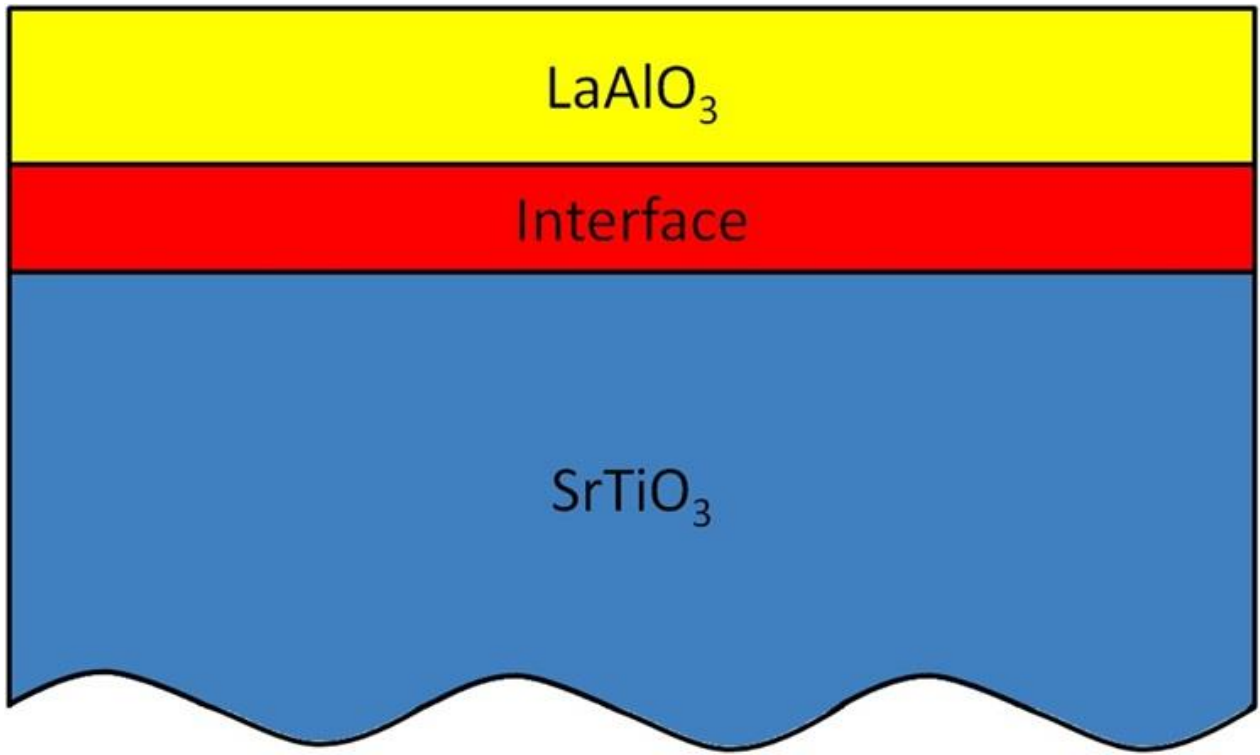

**Supplementary Figure 2: Multilayer consideration of conducting LaAlO$_3$/SrTiO$_3$.** The three constituent layers of LaAlO$_3$/SrTiO$_3$ are shown: LaAlO$_3$ film layer on top, bulk SrTiO$_3$ substrate at the bottom, and an interface layer sandwiched in between, representing the 2DEG of the conducting samples.

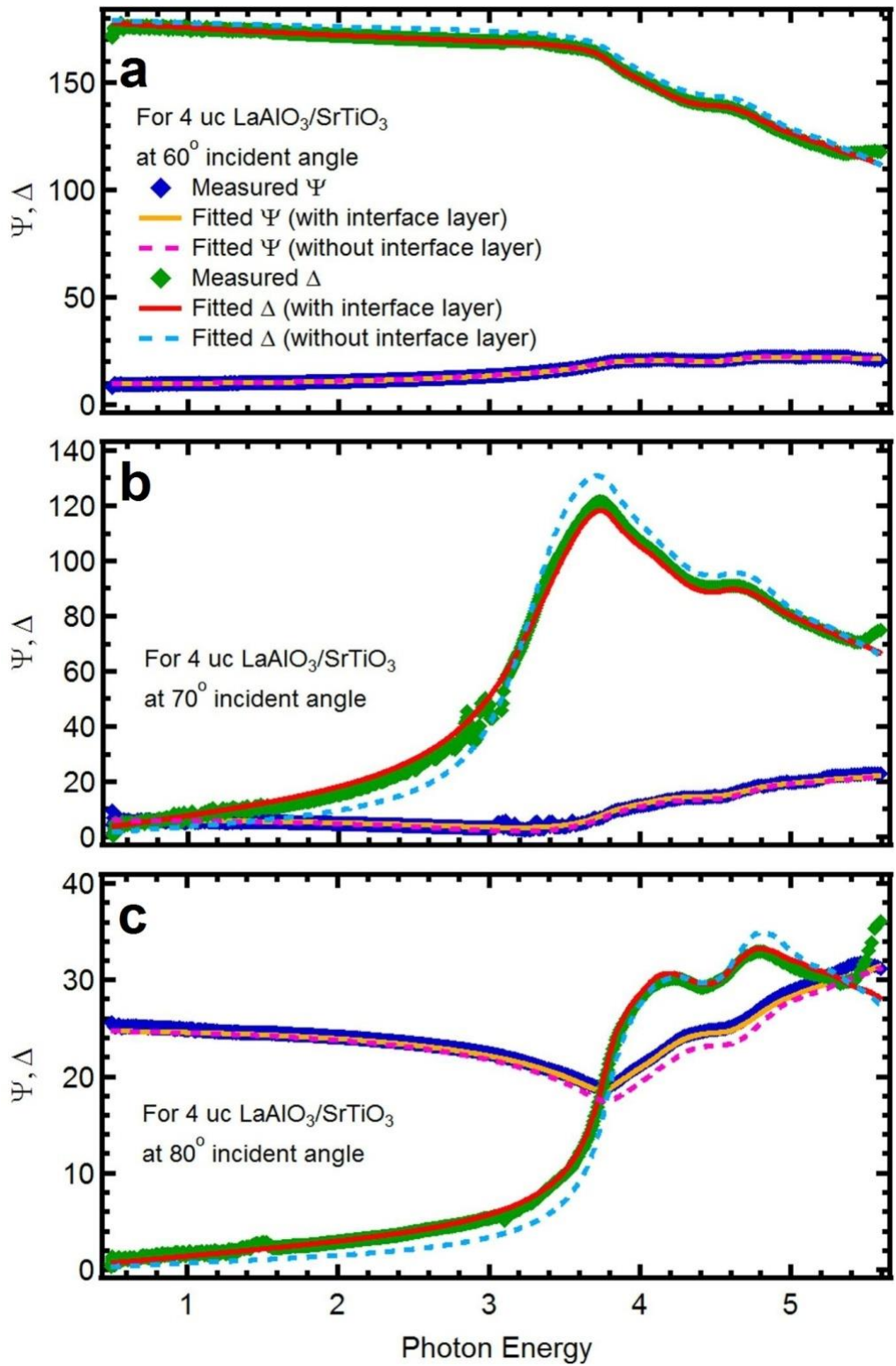

**Supplementary Figure 3: Fitted Ψ and Δ of 4 uc LaAlO$_3$/SrTiO$_3$ as compared to their measured values.** (**a**) For 60° incident angle. (**b**) For 70° incident angle. (**c**) For 80° incident angle. The fitted values match the measured Ψ and Δ very well for all three incident angles, thus confirming the stability of the iteration process.

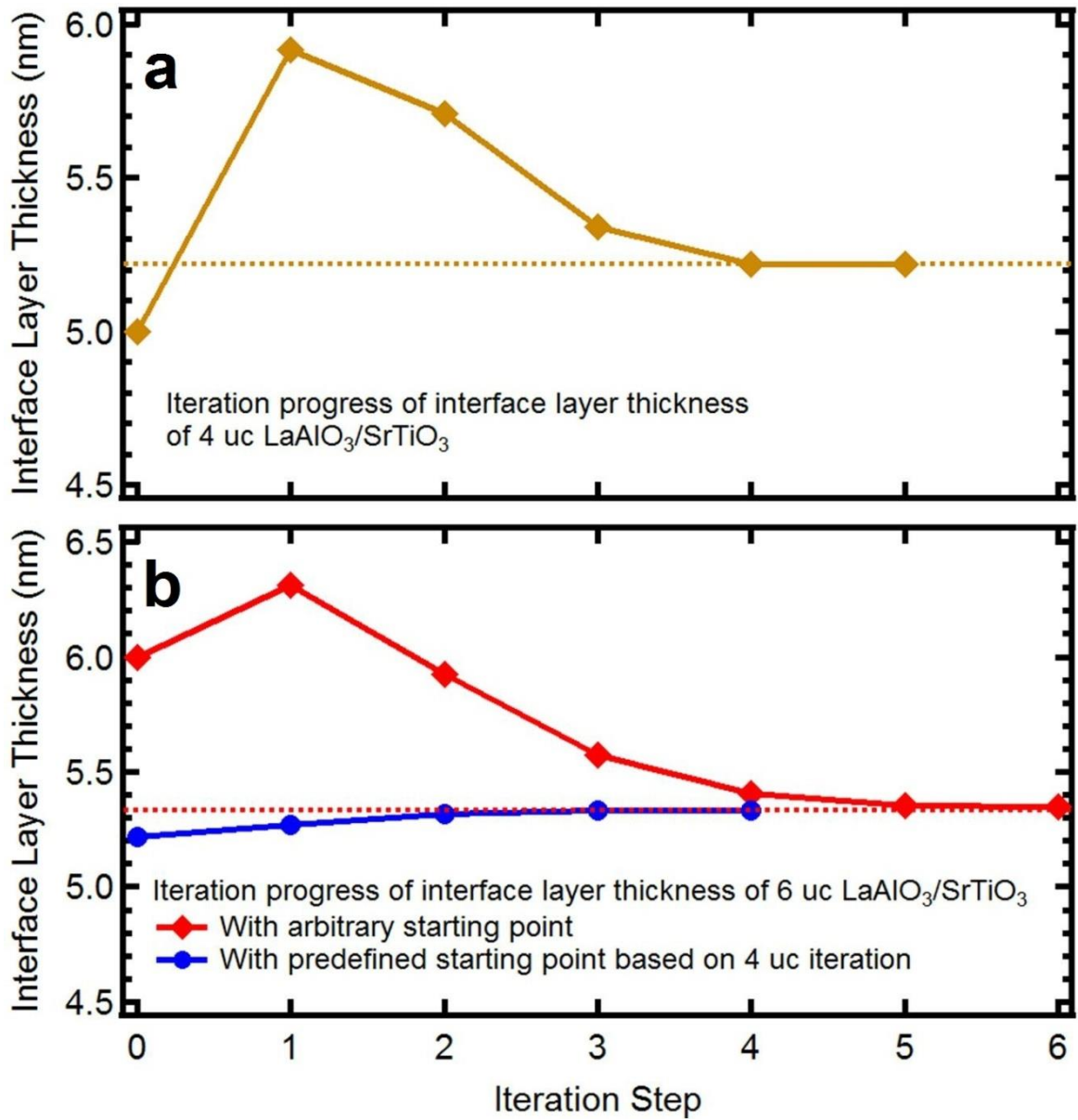

**Supplementary Figure 4: Iteration progress of the interface layer thickness.** (**a**) Iteration progress of the interface layer thickness for 4 uc LaAlO$_3$/SrTiO$_3$. (**b**) Iteration progress of the interface layer thickness for 6 uc LaAlO$_3$/SrTiO$_3$, showing the comparison between the two starting points.

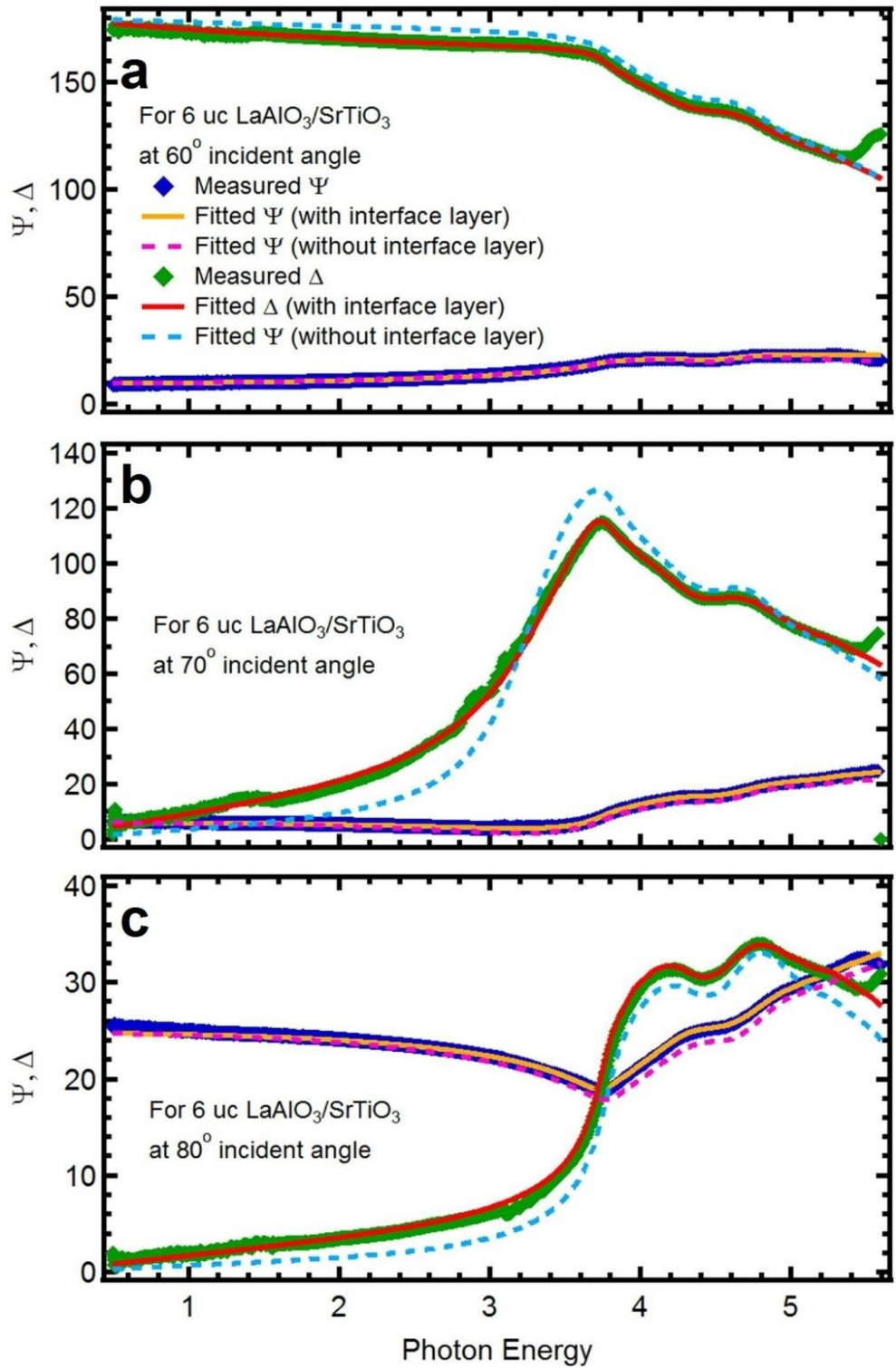

**Supplementary Figure 5: Fitted Ψ and Δ of 4 uc LaAlO$_3$/SrTiO$_3$ as compared to their measured values.** (**a**) For 60° incident angle. (**b**) For 70° incident angle. (**c**) For 80° incident angle. The fitted values match the measured Ψ and Δ very well for all three incident angles, thus confirming the stability of the iteration process.

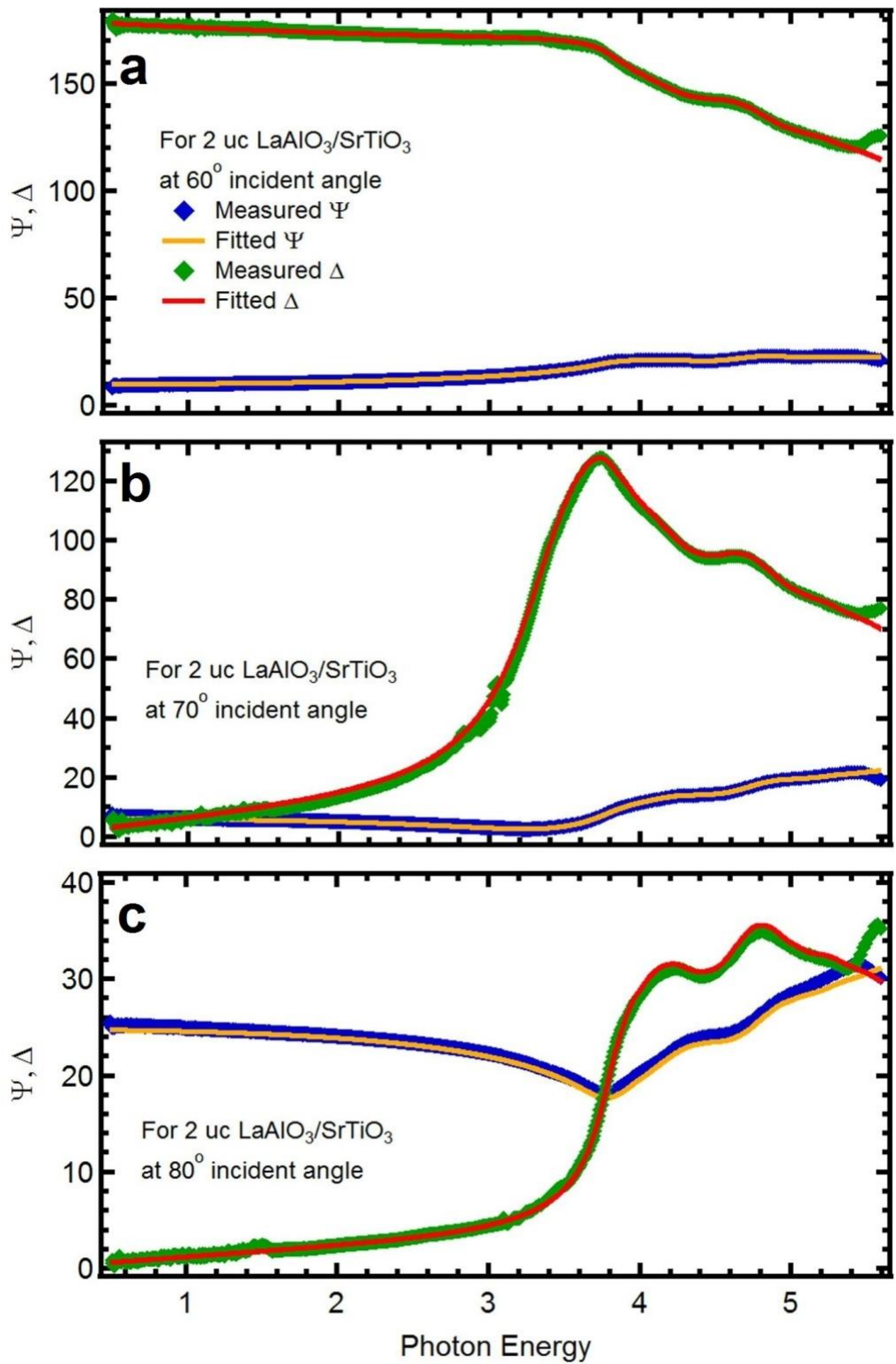

**Supplementary Figure 6: Fitted Ψ and Δ of 4 uc LaAlO₃/SrTiO₃ as compared to their measured values.** (**a**) For 60° incident angle. (**b**) For 70° incident angle. (**c**) For 80° incident angle. The fitted values match the measured Ψ and Δ very well for all three incident angles, thus confirming the stability of the iteration process.

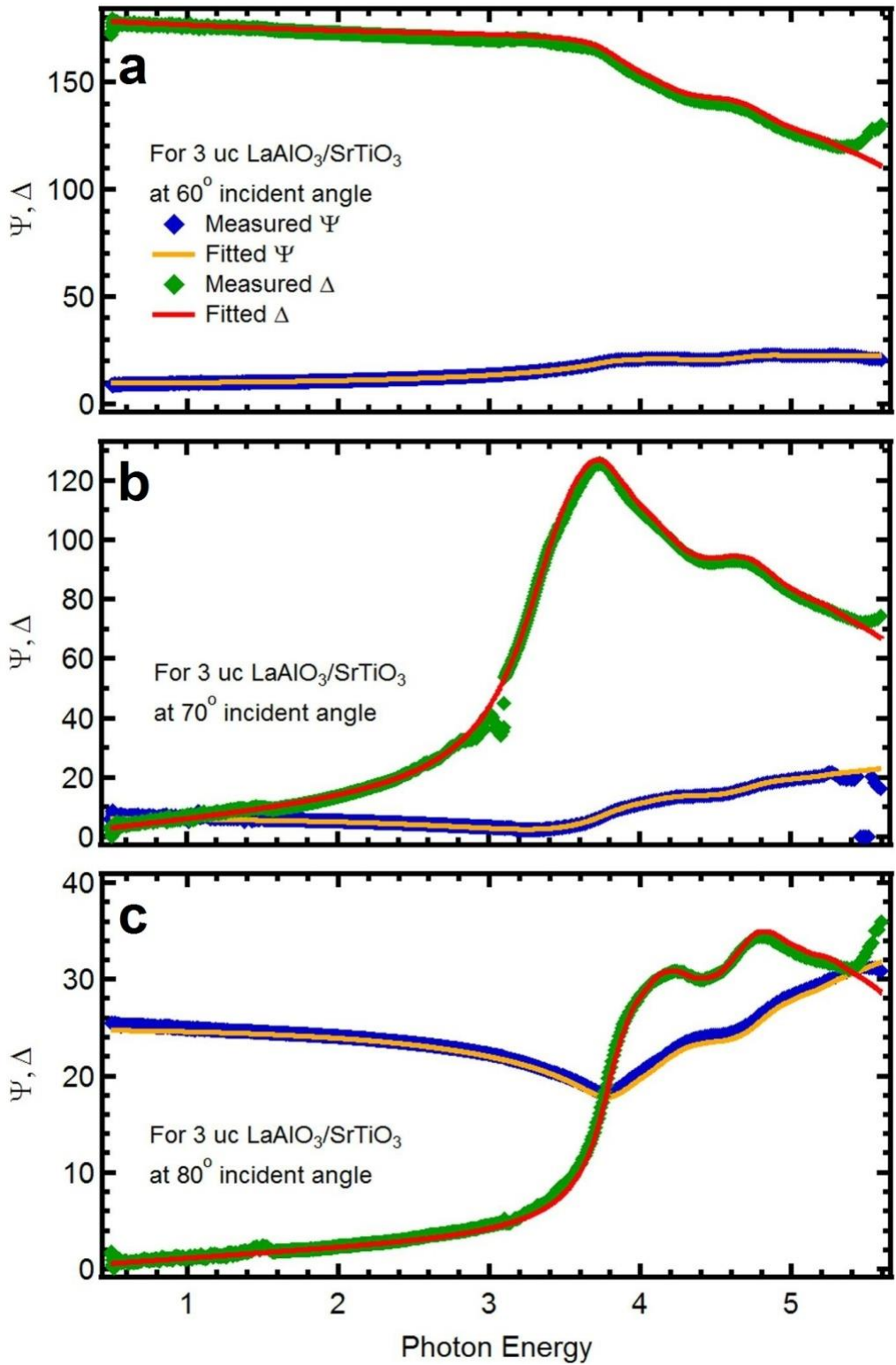

**Supplementary Figure 7| Fitted Ψ and Δ of 4 uc LaAlO₃/SrTiO₃ as compared to their measured values.** (**a**) For 60° incident angle. (**b**) For 70° incident angle. (**c**) For 80° incident angle. The fitted values match the measured Ψ and Δ very well for all three incident angles, thus confirming the stability of the iteration process.

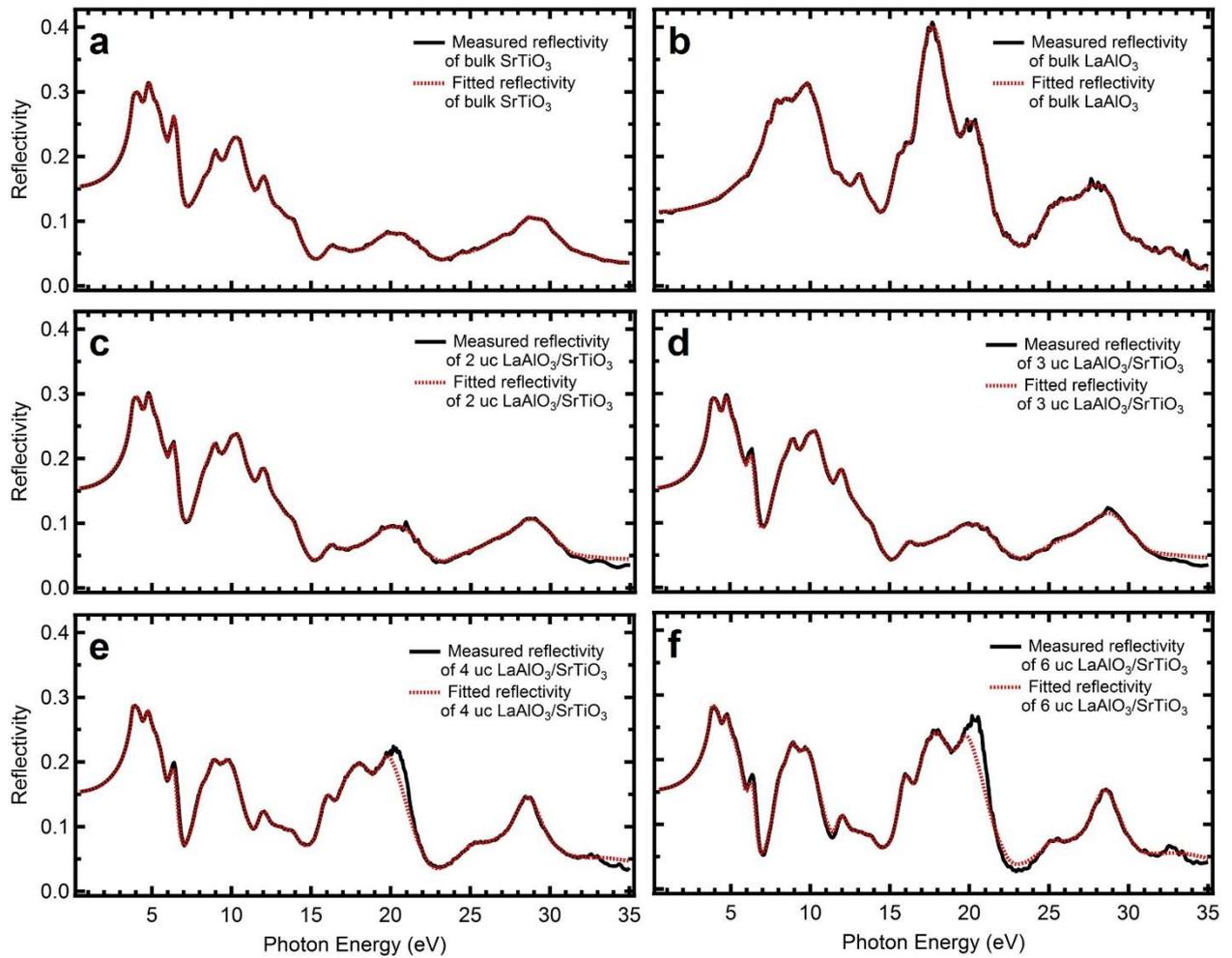

**Supplementary Figure 8: Fitted reflectivity of each sample as compared to its experimentally measured values.** (**a**) Fitted and measured reflectivity of bulk SrTiO$_3$. (**b**) Fitted and measured reflectivity of bulk LaAlO$_3$. (**c**) Fitted and measured reflectivity of 2 unit cells (uc) LaAlO$_3$/SrTiO$_3$. (**d**) Fitted and measured reflectivity of 3 uc LaAlO$_3$/SrTiO$_3$. (**e**) Fitted and measured reflectivity of 4 uc LaAlO$_3$/SrTiO$_3$. (**f**) Fitted and measured reflectivity of 6 uc LaAlO$_3$/SrTiO$_3$.